\documentclass[aps,prd,groupedaddress,showpacs,nofootinbib,amssymb,balancelastpage,preprintnumbers]{revtex4}
\usepackage[dvipdfmx]{graphicx}
\usepackage{graphicx,bm,color}
\usepackage{subfigure}
\usepackage{amsmath}
\usepackage{amssymb}
\usepackage{amsfonts}
\usepackage{cases}
\usepackage{cancel}
\usepackage{hyperref}
\usepackage{ulem}

\usepackage{here}
\usepackage{tikz}
\usetikzlibrary{matrix}

\begin{document}

\title{
New aspect of chiral $SU(2)$ and $U(1)$ axial breaking in QCD 
}

\author{Chuan-Xin Cui}\thanks{{\tt cuicx1618@mails.jlu.edu.cn}}
\affiliation{Center for Theoretical Physics and College of Physics, Jilin University, Changchun, 130012,
China}

\author{Mamiya Kawaguchi}\thanks{{\tt mamiya@ucas.ac.cn}} 
      \affiliation{ 
School of Nuclear Science and Technology, University of Chinese Academy of Sciences, Beijing 100049, China
}

\author{Jin-Yang Li}\thanks{{\tt lijy1118@mails.jlu.edu.cn}}
\affiliation{Center for Theoretical Physics and College of Physics, Jilin University, Changchun, 130012,
China}

\author{Shinya Matsuzaki}\thanks{{\tt synya@jlu.edu.cn}}
\affiliation{Center for Theoretical Physics and College of Physics, Jilin University, Changchun, 130012,
China}

\author{Akio Tomiya}\thanks{{\tt akio@yukawa.kyoto-u.ac.jp}}
\affiliation{
Department of Information Technology, International Professional University of Technology in Osaka, 3-3-1 Umeda, Kita-Ku, Osaka, 530-0001, Japan
}

\begin{abstract}
Violation of the $U(1)$ axial symmetry in QCD is 
\textcolor{black}{stricter}
than the chiral $SU(2)$ breaking, simply because of the presence of the quantum axial anomaly. If the QCD gauge coupling is sent to zero \textcolor{black}{(the asymptotic free limit, where the $U(1)$ axial anomaly does not exist),} the strength of the $U(1)$ axial breaking coincides with that of the chiral $SU(2)$ breaking, which we shall in short call an axial-chiral coincidence. This coincidence is {\it trivial} since QCD then becomes a non-interacting theory. Actually, there exists another limit in the QCD parameter space, where an axial-chiral coincidence occurs even with nonzero QCD gauge coupling, that can be dubbed a {\it nontrivial} coincidence: it is the case with the massive light quarks  $(m_l\neq 0)$ and the massless strange quark ($m_s=0$), due to the flavor-singlet nature of the topological susceptibility. This coincidence is robust and \textcolor{black}{tied to} the anomalous chiral Ward-Takahashi identity, which is operative even at hot QCD. This implies that the chiral $SU(2)$ symmetry \textcolor{black}{is restored simultaneously} with the $U(1)$ axial symmetry at high temperatures. This simultaneous restoration is independent of $m_l (\neq 0)$, hence is irrespective to the order of the chiral phase transition. In this \textcolor{black}{paper}, we discuss how the real-life QCD can be evolved from the nontrivial chiral-axial coincidence limit, by working on a Nambu-Jona-Lasinio model with the $U(1)$ axial anomaly contribution properly incorporated. 
\textcolor{black}{
It is shown that at high temperatures the large differences between the restorations of the chiral $SU(2)$ symmetry and the $U(1)$ axial symmetry for two light quarks and a sufficiently large current mass for the strange quark is induced by a significant interference of the topological susceptibility. 
} Thus the deviation from the nontrivial coincidence,  which is monitored by the strange quark mass controlling the topological susceptibility, provides a new way \textcolor{black}{of understanding the chiral $SU(2)$ and $U(1)$ axial breaking in QCD.} 

\end{abstract}

\maketitle

\section{Introduction}


\textcolor{black}{
The $U(1)$ axial symmetry in QCD (denoted as $U(1)_A$)} is explicitly broken by gluonic quantum corrections, called the $U(1)_A$ anomaly (or the axial anomaly), as well as by the current quark masses. 
\textcolor{black}{
The axial anomaly survives even in the limit of massless quarks.
}  
Thereby the $U(1)_A$ anomaly is 
anticipated to significantly interfere with 
the spontaneous breaking of
\textcolor{black}{
the $SU(2)$ axial (referred to as chiral $SU(2)$) symmetry
} 
in the nonperturbative QCD vacuum, i.e., 
the quark condensate, 
hence \textcolor{black}{affecting} the chiral phase transition in hot QCD.

As a pioneer work 
based on the renormalization group running in a chiral effective model,  
Pisarski and Wilczek \cite{Pisarski:1983ms}
pointed out that
\textcolor{black}{
the $U(1)_A$ anomaly, 
as well as the number of the quark-flavors, affect the order of the chiral phase transition in massless QCD. 
}
\textcolor{black}{Based on this}, 
the order of the chiral phase transition depending on the quark flavors   
has extensively been explored \textcolor{black}{in lattice QCD} 
in terms of the universality class~\cite{Ejiri:2009ac, Kaczmarek:2011zz,Engels:2011km,Burger:2011zc,HotQCD:2018pds,HotQCD:2019xnw}.  
\textcolor{black}{
The chiral phase transition is mapped onto a phase diagram in the quark mass plance, called the Columbia plot
} \cite{Brown:1990ev}. 
However, the anomalous $U(1)_A$ \textcolor{black}{contribution to} the whole phase diagram has not been clarified yet. 

Though there is no definite order parameter, 
the strength of the $U(1)_A$ breaking can be indicated 
by meson \textcolor{black}{correlation functions $\chi_{\rm meson}$ (susceptibility functions)} 
for the $U(1)_A$ \textcolor{black}{partners within the $U(2)$ meson quartet}:  $\chi_\sigma$ and $\chi_\eta$ for $(\sigma, \eta)$ mesons, 
and $\chi_\pi$ and $\chi_\delta$ for $(\pi, \delta)$ mesons. 
Those meson susceptibility functions are transformed 
also by the chiral \textcolor{black}{$SU(2)$ 
}  
rotation, like $(\chi_\sigma, \chi_\eta) \leftrightarrow (\chi_\pi, \chi_\delta)$ 
[See also Eq.(\ref{chi:transform})].
However, 
there exists a discrepancy between the meson susceptibility functions for the chiral symmetry,  
\textcolor{black}{$\chi_{\sigma}\nsim \chi_\pi$ and  $\chi_\delta\nsim \chi_\eta$}, and the axial symmetry, 
\textcolor{black}{$\chi_{\sigma}\nsim \chi_\eta$ and $\chi_{\delta}\nsim \chi_\pi$}, 
due to the spontaneous chiral symmetry breaking entangled with the $U(1)_A$ anomaly.
Hence,
the meson susceptibility functions 
cannot generically disentangle 
the $U(1)_A$ anomaly contribution \textcolor{black}{
from the contribution due to spontaneous chiral breaking.
}

With the increase of the temperature,
the meson susceptibility functions for the chiral partners 
turn to be degenerate, as a consequence of the chiral \textcolor{black}{
restoration, $\chi_\sigma\sim\chi_\pi$ and $\chi_\delta\sim\chi_\eta$, where the spontaneous breaking strength is separated out.
} 
Similarly, 
the approximate axial restoration can be seen from the degeneracy of the axial partner in the meson susceptibility functions: $\chi_\sigma \sim \chi_\eta$ and $\chi_\pi \sim \chi_\delta$. 
\textcolor{black}{Those degeneracies} has been observed in the lattice QCD simulations \textcolor{black}{with $2+1$ flavors at physical quark masses} \cite{Buchoff:2013nra,Bhattacharya:2014ara}.
In this context, 
\textcolor{black}{lattice QCD} 
has also shown that the chiral symmetry tends to be restored faster than the $U(1)_A$ symmetry at around the (pseudo)critical temperature \cite{Buchoff:2013nra, Bhattacharya:2014ara}. 
This discrepancy between the chiral and axial symmetry restorations  
would be caused by the existence  of the $U(1)_A$ anomaly, and this might be  the role of $U(1)_A$ \textcolor{black}{anomaly in the} spontaneous breaking of the chiral symmetry. Furthermore, \textcolor{black}{even if the light quarks become} massless and the strange quark mass takes the physical value, the $U(1)_A$ anomaly contribution remains \textcolor{black}{manifest} in the meson susceptibility functions at high temperatures~\cite{Ding:2020xlj}.

However, in contrast to the $2+1$ flavor QCD, it has been discussed 
that within the two-flavor QCD at the chiral limit, the $U(1)_A$ anomaly does not affect the symmetry restoration \cite{Cohen:1996ng,Aoki:2012yj,Cossu:2013uua,Tomiya:2016jwr,Aoki:2020noz}.
\textcolor{black}{
Therefore, it is still unclear how the $U(1)_A$ anomaly could contribute to the chiral breaking in terms of quark-flavor and mass dependencies.
} 



Another important aspect regarding the $U(1)_A$ anomaly that one should note is the close correlation with the transition rate of the \textcolor{black}{topological charge of the vacuum}, i.e., 
the topological susceptibility $\chi_{\rm top}$. 
Reflecting the flavor singlet nature of the QCD $\theta$-vacuum~\cite{Baluni:1978rf,Kim:1986ax}, 
$\chi_{\rm top}$ is given as the sum of the quark condensates coupled to the current quark masses and pseudoscalar susceptibilities \cite{Kawaguchi:2020qvg}:
$\chi_{\rm top}$ vanishes if either of quarks get massless. 
It is interesting to note that 
$\chi_{\rm top}$ can be rewritten 
as the meson susceptibility functions $\chi_\pi, \chi_{\delta (\sigma)}, $ and $\chi_\eta$, 
by using the anomalous Ward-Takahashi identity for 
\textcolor{black}{
the chiral symmetry with three quark flavors}~\cite{GomezNicola:2016ssy,GomezNicola:2017bhm,GomezNicola:2018pbx,Kawaguchi:2020qvg} 
as 
\textcolor{black}{$\chi_\eta-\chi_{\delta/\sigma}=\chi_\pi-\chi_{\delta/\sigma}+4\chi_{\rm top}/m_l^2$}
 ~\cite{Cui:2021bqf}, 
where $m_l$ denotes the current mass of the 
up and down quarks.
Note that the susceptibility difference \textcolor{black}{on both sides of this identity} plays the role of indicators of the breaking 
strength of the chiral or axial symmetry. 
\textcolor{black}{This identity shows that 
$\chi_{\rm top}$ is also important to explain} 
the chiral and axial symmetry restorations through the meson susceptibility functions.

The transparent link of $\chi_{\rm top}$ with the chiral or axial breaking can be observed in another way: 
it is seen through the Veneziano-Witten formula based on the current algebra assumed for $U(1)_A$ symmetry
~\cite{Witten:1979vv,Veneziano:1979ec}, 
$m_\eta^2 f_\eta^2 \sim \chi_{\rm top, g}$ 
where $\chi_{\rm top, g}$ is the contribution of pure gluonic diagrams.  
Though it is formulated 
in the large $N_c$ approximation for massless quarks 
(with $N_c$ being the number of colors), 
the aspect of the formula makes transparent that 
at high temperatures, 
the smooth decrease of the
topological susceptibility  is a combined effect of melting of the chiral condensate (supposed to be scaled along with $f_\eta$) and
suppression of the anomalous contribution to the mass of the isosinglet $\eta$, $m_\eta$ (in the heavy quark limit).


\textcolor{black}{Anyhow}, it is true that if the gluonic $U(1)_A$ anomaly is removed, the strength of the chiral symmetry breaking \textcolor{black}{will coincide} with the strength of the axial symmetry breaking in the meson susceptibility functions.
This corresponds to merely a trivial limit of QCD (with the gauge coupling sent to \textcolor{black}{zero}), 
\textcolor{black}{in which the understanding of the symmetry restoration will obviously become transparent}.  
However, the gluonic $U(1)_A$ anomaly is essential in the underlying QCD as an interacting gauge theory, 
so that    
its contribution would inevitably produce the intricate restoration phenomena involving contamination 
of the chiral and axial breaking, as mentioned above.

\textcolor{black}{Therefore, in this paper, we focus} on another limit where a nontrivial axial-chiral coincidence occurs even with nonzero QCD gauge coupling. 
The case \textcolor{black}{
we consider here is the one of massive light quarks ($m_l\ne 0$) and amassless strange quark ($m_s=0)$,
} 
due to the flavor-singlet nature of the topological susceptibility. 
\textcolor{black}{
Following a robust anomalous chiral Ward-Takahashi identity, this nontrivial coincidence is valid even at finite temperatures,
} so that the chiral symmetry gets restored simultaneously with the $U(1)_A$ 
symmetry at high temperatures, no matter what order of the chiral phase \textcolor{black}{
transition is performed
}.

\textcolor{black}{Usually, the topological susceptibility is used} as a probe for
the effective restoration of the $U(1)$ axial symmetry when the chiral $SU(2)$ symmetry is restored at high \textcolor{black}{temperatures. This is the way the differences in} the restorations of chiral $SU(2)$ and $U(1)$ axial symmetries can be monitored by the temperature dependence of the topological susceptibility.
However, this ordinary approach suffers 
\textcolor{black}{
from the practical difficulty to access
} the origin of the split 
without ambiguity, because the chiral $SU(2)$ symmetry breaking is highly contaminated 
with the $U(1)$ axial symmetry breaking even at the beginning, at the vacuum.

In contrast to the ordinary approach under the temperature control, the nontrivial coincidence discussed in the present paper provides a new approach: it is the strange quark mass that controls the strengths of
the chiral \textcolor{black}{$SU(2)$} symmetry breaking and the $U(1)$ axial symmetry breaking, and those strengths coincide in the limit $m_s=0$.

Thus, the gap between the breaking strengths of the chiral $SU(2)$ and $U(1)$ axial symmetries
 is handled by the strange quark mass, \textcolor{black}{
as it is the case for the $U(1)_A$ anomaly 
 }. 
 This can be thought of as an alternative to large-$N_c$ QCD a la Veneziano and Witten, \textcolor{black}{
 as our approach uses fixed $N_c=3$ and varies the current quark masses.}

To understand the real-life QCD departing from the nontrivial coincidence limit, 
we employ the Nambu-Jona-Lasinio (NJL) model. 
We first confirm that the NJL model surely provides the nontrivial coincidence at the massless limit of the strange quark.  
Once the strange quark gets massive, the strange quark mass handles the deviation from the nontrivial coincidence.
\textcolor{black}{We explain how the large differences} in the restorations of the \textcolor{black}{chiral $SU(2)$ 
} symmetry and the $U(1)_A$ symmetry  
\textcolor{black}{
for $2+1$ flavors with physical quark masses is generated by a
} 
sufficiently large current mass of the strange quark through
the significant interference of the topological susceptibility.

\textcolor{black}{
The deviation from the nontrivial coincidence as monitored by the strange quark mass by controlling
}
the topological susceptibility provides a new way of understanding 
of the chiral and axial breaking in QCD, 
\textcolor{black}{
seen on the graphical description given by the Columbia plot.
}


\section{
Coincidence between chiral and axial symmetry breaking
}

As noted in the Introduction, 
the meson susceptibility function plays \textcolor{black}{
the role of an indicator
} for the chiral \textcolor{black}{$SU(2)$} symmetry breaking and the $U(1)$ axial symmetry breaking.  
The $U(1)_A$ anomaly potentially produces a difference between the strength of the chiral symmetry breaking and the axial symmetry breaking 
in the meson susceptibility functions
\textcolor{black}{of the vacuum}. 
In this section, we show \textcolor{black}{
that even in keeping a nonzero
} $U(1)_A$ anomaly,  
there generically exists 
the nontrivial coincidence between the chiral and axial symmetry breaking \textcolor{black}{in} QCD for $N_f=2+1$ flavors.

\subsection{Chiral and axial symmetry in meson susceptibility functions}
We begin by introducing the scalar- and pseudoscalar-meson susceptibilities. 
The pion susceptibility $\chi_{\pi}$, the $\eta$-meson susceptibility $\chi_\eta$, the $\delta$-meson susceptibility $\chi_\delta$
(also known as $a_0$ meson), 
and 
the $\sigma$-meson susceptibility $\chi_\sigma$
are defined respectively as  
\begin{eqnarray}
	\chi_{\pi}
	&=&\int_T d^4 x 
	\left[ 
	\langle (i \bar u(0) \gamma_5  u(0))( i\bar u(x) \gamma_5 u(x))\rangle_{\rm conn}
+ \langle ( i\bar d(0)  \gamma_5 d(0))(i \bar d(x) \gamma_5 d(x))\rangle_{\rm conn}
\right]
\,, \nonumber\\
\chi_{\eta} 
    &=&\int_T d^4 x 
	\left[ 
	\langle (i\bar u(0) \gamma _5  u(0))
	(i\bar u(x) \gamma _5 u(x))\rangle 
	+
	\langle (i\bar d(0) \gamma _5  d(0))
	(i\bar d(x) \gamma _5 d(x))\rangle
	+
	2 
	\langle (i\bar u(0) \gamma _5  u(0))
	(i\bar d(x) \gamma _5 d(x))\rangle
	\right] 
	\, , \nonumber\\
	\chi_{\delta} 
	&=&\int_T d^4 x 
	\left[ 
	\langle ( \bar u(0)  u(0))( \bar u(x) u(x))\rangle_{\rm conn}
+ \langle ( \bar d(0)  d(0))( \bar d(x) d(x))\rangle_{\rm conn} 
\right]
\, ,\nonumber\\
 \chi_\sigma&=&\int_T d^4 x 
	\left[ 
	\langle (\bar u(0)   u(0))
	(\bar u(x)  u(x))\rangle 
	+
	\langle (\bar d(0)   d(0))
	(\bar d(x)  d(x))\rangle
	+
	2 
	\langle (\bar u(0)  u(0))
	(\bar d(x)  d(x))\rangle
	\right] 
	\, , \label{chi-meson--def}
\end{eqnarray}
where
$u$ and $d$ are the up- and down-quark fields;  
$\langle \cdots \rangle_{\rm conn}$ represents the connected-graph part of the correlation
function;
$\int_T d^4 x \equiv \int^{1/T}_0 d \tau \int d^3 x$ with the imaginary time $\tau = i x_0$.
Under the \textcolor{black}{chiral $SU(2)$} rotation
and the $U(1)_A$ rotation, 
the meson susceptibility functions can be exchanged with each other:
\begin{eqnarray}
\mbox{
Chiral $SU(2)$
rotation}:
&&
\begin{cases}
\chi_\delta\leftrightarrow \chi_\eta\\
\chi_\sigma\leftrightarrow
\chi_\pi
\end{cases}
,
\nonumber\\
\mbox{$U(1)_A$ rotation}
&&
\begin{cases}
\chi_\delta\leftrightarrow \chi_\pi\\
\chi_\sigma\leftrightarrow
\chi_\eta
\end{cases}
. \label{chi:transform}
\end{eqnarray}
\textcolor{black}{For the convenience of the reader}, we also provide the following cartoon
\textcolor{black}{in order to visualize} the chiral \textcolor{black}{$SU(2)$} and $U(1)_A$ \textcolor{black}{transformations for the meson susceptibility functions,}  
\begin{center}
\begin{tikzpicture}
  \matrix (m) [matrix of math nodes,row sep=4em,column sep=5em,minimum width=3em]
  {
     \chi_{\pi} & \chi_{\sigma} \\
     \chi_{\delta} & \chi_{\eta} \\};
  \path[-stealth]
    (m-1-1) edge node [midway,left] {$U(1)_A$} (m-2-1)
            edge  node [above] {SU(2)} (m-1-2)
    (m-2-1) edge node [below] {SU(2)} (m-2-2)
    (m-1-2) edge node [right] {$U(1)_A$} (m-2-2)
    (m-2-1) edge node [midway,left] { } (m-1-1)
    (m-1-2) edge node [midway,left] { } (m-1-1)
    (m-2-2) edge node [midway,left] { } (m-1-2)
    (m-2-2) edge node [midway,left] { } (m-2-1);
\end{tikzpicture}
\end{center}

Since the meson susceptibility functions are linked with each other via the chiral symmetry and the axial symmetry,
the chiral and axial partners 
become (approximately)
degenerate, respectively, 
at the restoration limits of the chiral \textcolor{black}{$SU(2)$} symmetry and the $U(1)_A$ symmetry:
\begin{eqnarray}
\mbox{
Chiral $SU(2)$
symmetric limit}:
&&
\begin{cases}
\chi_{\eta-\delta}=\chi_{\eta}-\chi_\delta\to 0\\
\chi_{\pi-\sigma}=
\chi_\pi-\chi_\sigma\to0
\end{cases}
,
\nonumber\\
\mbox{$U(1)_A$ symmetric limit}:&&
\begin{cases}
\chi_{\pi-\delta}
=
\chi_{\pi} -\chi_{\delta}\to 0
\\
\chi_{\eta-\sigma}
=
\chi_{\eta}-\chi_\sigma\to 0
\end{cases}
.
\end{eqnarray}
Note that in the chiral limit, 
we encounter the infrared divergence in 
the pseudoscalar-meson susceptibilities 
due to the emergence of the exactly massless Nambu-Goldstone bosons. 
The nonzero light quark mass thus \textcolor{black}{
plays the role of a regulator for the infrared divergence, making them well-defined.
} 
\textcolor{black}{The differences of the susceptibility functions of the mesons} forming the chiral and axial partners,   
$\chi_{\eta-\delta}$,
$\chi_{\pi-\sigma}$, 
$\chi_{\pi-\delta}$,
and $\chi_{\eta-\sigma}$, 
can safely serve as well-defined indicators for the symmetry breaking of  $SU(2)_L\times SU(2)_R$ and $U(1)_A$.

The pseudoscalar susceptibility functions
have close correlations with the quark condensates through the anomalous Ward-Takahashi identities for 
\textcolor{black}{
the chiral symmetry with the three quark flaovrs
} \cite{GomezNicola:2016ssy,Kawaguchi:2020qvg}: 
\begin{align} 
\langle \bar uu \rangle +\langle \bar dd \rangle 
 &= - m_l \chi_\pi 
\,, \notag\\  
\langle \bar uu \rangle +\langle \bar dd \rangle
+ 4 \langle \bar s s \rangle
& = 
- \left[ m_l 
\chi_\eta
- 2 (m_s + m_l)\left(\chi_{P}^{us}+\chi_{P}^{ds} \right)
+ 4 m_s \chi_P^{ss} \right] 
\,, \notag\\ 
\langle \bar uu \rangle +\langle \bar dd \rangle
-2  \langle \bar s s \rangle
& = 
- \left[ m_l 
\chi_\eta
+  (m_l - 2 m_s)\left(\chi_{P}^{us}+\chi_{P}^{ds} \right) 
-  2 m_s \chi_P^{ss} \right] 
		\label{chipi}\,, 
	\end{align} 
where
$m_l=m_u=m_d$ is the isospin-symmetric mass for
the up- and down quarks, 
$m_s$ is the strange quark mass, 
and
the pseudoscalar susceptibilities 
$\chi_{P}^{f_1f_2}$, 
are defined as 
\begin{align} 
        \chi_P^{f_1f_2} &=\int_T  d^4 x \langle (\bar q_{f_1}(0) i \gamma _5 q_{f_1}(0))(\bar q_{f_2}(x) i \gamma _5 q_{f_2}(x))\rangle
\,, \qquad {\rm for} \quad {q_{_{f_{1,2}}} = u,d,s}.
\label{f_chi}
\end{align}
\textcolor{black}{
In addition, we also define the scalar susceptibilities $\chi_S^{f_1f_2}$,
\begin{align} 
        \chi_S^{f_1f_2} &=\int_T  d^4 x \langle (\bar q_{f_1}(0) q_{f_1}(0))(\bar q_{f_2}(x)  q_{f_2}(x))\rangle.
\label{f_chi_S}
\end{align}
}

 Due to the the anomalous Ward-Takahashi identities in  Eq.~(\ref{chipi}),
the behavior of the quark condensates 
\textcolor{black}{close to}
the chiral phase transition is directly reflected in that of the pseudoscalar susceptibility functions. 

\subsection{
Trivial coincidence between chiral 
and axial symmetry breaking
}\label{triv_coin_sub}

In the previous subsection, we have shown that 
$\chi_{\eta-\delta}$ or $\chi_{\pi-\sigma}$ 
($\chi_{\pi-\delta}$ or $\chi_{\eta-\sigma}$)
serves as an indicator for the strength of the chiral (axial) symmetry breaking.
The symmetry breaking in the meson susceptibility functions has 
been studied in a chiral-effective model approach \cite{GomezNicola:2016ssy,Cui:2021bqf}
and 
the first-principle calculations of lattice QCD \cite{Buchoff:2013nra,GomezNicola:2017bhm,Bhattacharya:2014ara,Cohen:1996ng,Aoki:2012yj,GomezNicola:2018pbx}. 
It is known that there exists a difference 
 between the indicators for the chiral $SU(2)_L\times SU(2)_R$ symmetry breaking and the $U(1)_A$ symmetry breaking at the vacuum: 
\begin{eqnarray}
\begin{cases}
\chi_{\eta-\delta}\neq \chi_{\pi-\delta}\\
\chi_{\pi-\sigma}\neq
\chi_{\eta-\sigma}
\end{cases}.
\label{discrepancy}
\end{eqnarray}
This discrepancy 
is originated from the anomalous current conservation laws for 
the $U(1)_A$ symmetry.
In the underlying QCD, 
the chiral current $j^{a\mu}_{A}$ and the axial current $j^{\mu}_{A}$ follow the following 
anomalous conservation laws: 
\begin{eqnarray}
\partial_\mu j_A^{a\mu }&=&
i\bar q\left\{\bm m, \frac{\tau^a}{2}   \right\}\gamma_5 q,
\nonumber\\
\partial_\mu j^\mu_A&=&2i \bar q {\bm m}\gamma_5 q
+N_f\frac{g^2}{32\pi^2}\epsilon^{\mu\nu\rho\sigma}  F _{\mu\nu}^a  F^a_{\rho\sigma},
\label{current_cons_law_QCD}
\end{eqnarray}
where 
$q$ is the $SU(3)$-flavor triplet-quark field $q=(u,d,s)^T$;  
the chiral current  and the axial current are defined as 
$j^{a\mu}_A=\bar q \gamma_\mu \gamma_5 \frac{\tau_a}{2}q$ and $j^\mu_A=\bar q \gamma_5 \gamma^\mu q$, respectively;   
$\tau^a (a=1,2,3)$ generate an $SU(2)$ subalgebra of Gell-Mann matrices;
${\bm m}$ denotes the mass matrix ${\bm m}={\rm diag}(m_u,m_d,m_s)$;  
$F_{\mu\nu}^a$ denotes the field strength of the gluon field; $g$ stands for the QCD coupling constant.
At the Lagrangian level,
the chiral \textcolor{black}{$SU(2)$} symmetry and the $U(1)_A$ symmetry are explicitly broken by the current quark mass terms, so that the chiral current and the axial current get anomalous parts from the quark mass terms in Eq.~(\ref{current_cons_law_QCD}).
These anomalous parts of the quark masses can be tuned to vanish by taking the chiral limit ${\bm m}=0 $. 

Looking at the QCD generating functional, one notices that
the gluonic quantum anomaly $g^2 \epsilon^{\mu\nu\rho\sigma}  F _{\mu\nu}^a  F^a_{\rho\sigma} $
arises in the axial current, but not in the chiral current.
The quantum correction only to the $U(1)_A$ symmetry
induces the sizable discrepancy between the chiral symmetry breaking and the axial symmetry breaking in Eq~(\ref{discrepancy}). 
In contrast to the anomalous term of 
the quark mass, the quantum gluonic anomaly cannot be eliminated
from Eq.~(\ref{current_cons_law_QCD})
by tuning the external parameters like the current quark masses.
If one tries to remove the gluonic quantum anomaly in Eq.~(\ref{current_cons_law_QCD}), the QCD coupling constant would be taken as $g=0$.
\textcolor{black}{The vanishing} quantum anomaly provides the coincidence between the strength of the chiral symmetry breaking and the axial symmetry breaking in the meson susceptibility functions\footnote{
In the free theory of quarks,
the chiral symmetry is not spontaneously broken, so that  
the $U(2)_L\times U(2)_R$ symmetry is realized in the meson susceptibility functions: 
$\chi_\pi=\chi_\sigma=\chi_\eta=\chi_\delta$ for $g=0$.
}: 
\begin{eqnarray}
\begin{cases}
\chi_{\eta-\delta}=
\chi_{\pi-\delta}\\
\chi_{\pi-\sigma}=
\chi_{\eta-\sigma}
\end{cases}
(\mbox{for $g=0$}).
\label{trivial_limit_QCD}
\end{eqnarray}
However, in this case
QCD obviously becomes a free theory and loses the
nontrivial  features driven by the interaction among quarks and gluons as quantum field theory. 

\subsection{
 Flavor singlet nature of topological susceptibility
}
\textcolor{black}{We will explain later} that the discrepancy between the meson susceptibility functions in Eq.~(\ref{discrepancy}) 
is actually responsible for nonzero topological susceptibility.
To make it better understood, 
in this subsection we give a brief review of the construction of the topological susceptibility \cite{Kawaguchi:2020qvg} and its flavor singlet nature. 
%


The topological susceptibility 
is a quantity to measure the
topological charge fluctuation of the QCD-$\theta$ vacuum, which
is defined as the curvature
of the $\theta$-dependent QCD vacuum energy $V(\theta)$ at $\theta =0$,
\begin{equation}
\chi_{\rm top}=-\int_T d^{4}x \frac{\delta^2 V(\theta)}{\delta \theta(x)\delta \theta(0) }\Bigg{|}_{\theta =0}\,. 
\label{chitop_f}
\end{equation}
$V(\theta)$ represents the effective potential of QCD, which includes  
the QCD $\theta$-term represented by the
flavor-singlet gluonic 
operator, $\theta g^2 \epsilon^{\mu\nu\rho\sigma}F_{\mu\nu}^a F^{a}_{\rho\sigma}$:
\begin{eqnarray}
V(\theta)&=&-\log Z_{\rm QCD}(\theta),
\nonumber\\
Z_{\rm QCD}(\theta)&=&\int [\Pi_f dq_f d\bar q_f] [dA]
\exp\Biggl[-\int_T d^4x \Biggl\{
\sum_{f=u,d,s}\Bigg(
\bar q^f_Li\gamma^\mu D_\mu q^f_L
+
\bar q^f_Ri\gamma^\mu D_\mu q^f_R
\notag\\ 
&& +\bar q^f_L m_f q^f_R+\bar q^f_R m_f q^f_L\Bigg)
+\frac{1}{4}(F_{\mu\nu}^a)^2 
+\frac{i\theta }{32\pi^2}g^2F_{\mu\nu}^a\tilde F_{\mu\nu}^a
\Biggl\}
\Biggl]
\label{effpotQCD}
\end{eqnarray}
with $Z_{\rm QCD}$ being the generating functional of QCD in Euclidean space.  
In Eq.(\ref{effpotQCD}) $q^f_{L(R)}$ denote the left- (right-) handed quark fields; the covariant derivative of the quark field is represented as $D_\mu$ involving the gluon fields $(A_\mu^a)$ and 
$F_{\mu\nu}^a$ is the field strength of the gluon field. From Eq.~(\ref{chitop_f}), 
the topological susceptibility $\chi_{\rm top}$ is directly given as 
\begin{eqnarray}
\chi_{\rm top}=
\int_T  d^4x \langle Q(x) Q(0) \rangle
\label{original_chitop}
\end{eqnarray}
with $Q=g^2/(32\pi^2)\, F_{\mu\nu}^a \tilde F^{a\mu\nu}$. 
Obviously the topological susceptibility in Eq.~(\ref{original_chitop}) \textcolor{black}{takes a flavor-independent form}. 

Under the $U(1)_A$ transformation with the rotation angles $\theta_{f=u,d,s}$,
the QCD-$\theta$ term
in the generating functional
is shifted by the $U(1)_A$ anomaly: 
\begin{eqnarray}
&&\int [\Pi_f dq_f d\bar q_f] [dA]
\exp\Biggl[-\int_T d^4x \Biggl\{
\sum_{f=u,d,s}\Bigg(
\bar q^f_Li\gamma^\mu D_\mu q^f_L
+
\bar q^f_Ri\gamma^\mu D_\mu q^f_R
\notag\\ 
&& 
+
\bar q^f_L m_f e^{i\theta_f} q^f_R+\bar q^f_R m_f e^{-i\theta_f} q^f_L\Bigg)
+\frac{1}{4}(F_{\mu\nu}^a)^2
+
\frac{i(\theta-\bar\theta)}{32\pi^2}
g^2F_{\mu\nu}^a\tilde F_{\mu\nu}^a
\Biggl\}
\Biggl],
\label{QCDGF_UAR}
\end{eqnarray}
where 
$\bar\theta=\sum_{f=u,d,s} \theta_f = \theta_u+\theta_d+\theta_s$.
The $\theta$-dependence of the topological operator $F \tilde{F}$ 
can be \textcolor{black}{transferred to} the quark mass terms by choosing the rotation angles as
\begin{eqnarray}
\bar \theta=\theta_u+\theta_d+\theta_s=\theta. 
\label{choicetheta_f}
\end{eqnarray}
\textcolor{black}{
Absorbing the $\theta$ dependence into the quark mass terms by this choice, the QCD $\theta$-term is removed from the generating functional.
}
However, currently the $\theta$-dependent quark mass term is not flavor universal, though 
the original QCD $\theta$-term is flavor independent.
Thus one should impose a flavor-singlet constraint on the axial rotation angles left in the quark mass sector as 
\cite{Baluni:1978rf,Kim:1986ax}
\begin{eqnarray}
m_u\theta_u=m_d\theta_d=m_s\theta_s 
\,, 
\label{singlet_con}
\end{eqnarray}
for $\theta \ll 1$, so that the $\theta$-dependent part of the quark mass term satisfies the flavor singlet nature:
\begin{eqnarray}
{\cal L}_{\rm QCD }^{(\theta)}
=
\sum_f\left(
\bar q^f_Li\gamma^\mu D_\mu q^f_L
+
\bar q^f_Ri\gamma^\mu D_\mu q^f_R
\right)
+
\bar q_L {\cal M}_\theta q_R+\bar q_R {\cal M}_\theta^\dagger q_L
+\frac{1}{4}(F_{\mu\nu}^a)^2
\,,
\label{TQCDlag}
\end{eqnarray}
where  ${\cal M}_\theta$ denotes the $\theta$-dependent quark matrix,
\begin{eqnarray}
{\cal M}_\theta=
{\rm diag}\left[m_u\exp\left(i\frac{\bar m}{m_u}\theta\right), m_d\exp\left(i\frac{\bar m}{m_d}\theta\right), m_s\exp\left(i\frac{\bar m}{m_s}\theta\right)\right],
\label{calM-theta}
\end{eqnarray} 
with $\bar m \equiv \left(\frac{1}{m_l}+\frac{1}{m_l}+\frac{1}{m_s} \right)^{-1}$.
We thus find \cite{Kawaguchi:2020qvg},
\begin{eqnarray}
		\chi_{\rm top} 
		&=&
		 \bar{m}^2 \left[ 
		\frac{\langle \bar {u}u \rangle}{m_l}  
		+\frac{\langle \bar {d}d \rangle}{m_l} 
		+\frac{\langle \bar {s}s \rangle}{m_s} 
		+  
		\chi_{P}^{uu}+\chi_{P}^{dd}
		+\chi_P^{ss} 
		+ 2 \chi_P^{ud} 
		+2\chi_{P}^{us}+
		2\chi_{P}^{ds} 
        \right] 
	\notag \\ 
	& =& \frac{1}{4} \left[ 
	m_l\left(\langle \bar{u} u \rangle +\langle \bar{d} d \rangle  \right)
	+ m_l^2 \chi_\eta
	\right] 
	= m_s \langle \bar{s}s \rangle + m_s^2 \chi_P^{ss} 
	\label{chitop_wfsc}
	\,. 
	\end{eqnarray}
It is important to note that
the topological susceptibility in Eq.~(\ref{chitop_wfsc}) vanishes if \textcolor{black}{either of the quarks} get massless,
\begin{eqnarray}
\chi_{\rm top}=0\;\;\;
(\mbox{for $m_l$ or $m_s$ = 0}), 
\label{FS_condition}
\end{eqnarray}
which reflects 
the flavor singlet nature of the QCD-$\theta$ vacuum. 

\subsection{Correlation between 
susceptibility functions} 

By combining the anomalous Ward-Takahashi identities in Eq.~(\ref{chipi}) with $\chi_{\rm top}$ in Eq.~(\ref{chitop_wfsc}),
the topological susceptibility can be also rewritten as 
\begin{eqnarray}
\chi_{\rm top}
&=&\frac{1}{4}m_l^2 (\chi_\eta-\chi_\pi)
\label{chitop_wfac}.
\end{eqnarray}
Inserting the scalar-meson susceptibility in Eq.~(\ref{chitop_wfac}),
we eventually obtain the crucial formula for understanding the QCD vacuum structure:
\begin{eqnarray}
\frac{1}{4}m_l^2 (\chi_{\eta-\delta}-\chi_{\pi-\delta})
&=&
\chi_{\rm top},
\nonumber
\\
\frac{1}{4}m_l^2 (\chi_{\eta-\sigma}-\chi_{\pi-\sigma})
&=&
\chi_{\rm top}.
\label{WI-def}
\end{eqnarray}

Of interest is that
the susceptibility functions for the chiral symmetry, the axial symmetry and the topological charge are merged \textcolor{black}{into a single equation}.
Therefore Eq.~(\ref{WI-def}) is valuable in considering the nontrivial correlation between the symmetry breaking and the topological feature in the susceptibility functions.
In particular, 
it is shown that
the difference  
between the indicator for the chiral symmetry breaking strength   $\chi_{\eta- \delta}$ ($\chi_{\pi-\sigma}$) and the indicator for the axial symmetry breaking strength
$ \chi_{\pi- \delta}$ ($\chi_{\eta-\sigma}$)
corresponds to the topological susceptibility.

\subsection{
Nontrivial  coincidence between chiral 
and axial symmetry breaking
}

As shown in the trivial limit of Eq.~(\ref{trivial_limit_QCD})
the strength of the chiral symmetry breaking certainly coincides with that of the axial one
 due to the absence of the gluonic quantum anomaly in the $U(1)_A$ symmetry.  
Once we include the quantum corrections in the QCD generating functional, 
it is inevitable that the $U(1)_A$ anomaly shows up in the axial current.  
Thus, one might think that there does not \textcolor{black}{exist the limit of the nontrivial} coincidence between the chiral and axial symmetry breaking strength in the nonperturbative QCD vacuum.
\textcolor{black}{However, paying attention} to the flavor singlet nature of the topological susceptibility,
\textcolor{black}{we can find a nontrivial coincidence while saving the gluonic $U(1)_A$ anomaly.}

From Eq.~(\ref{WI-def}), we note that the discrepancy between the strength of the chiral and axial symmetry breaking in the meson susceptibility functions
can be controlled by the topological susceptibility. 
This implies that 
the discrepancy can be tuned to be zero by taking the massless limit of the strange quark, 
due to the flavor singlet nature in  Eq.~(\ref{FS_condition}):
\begin{eqnarray}
\begin{cases}
\chi_{\eta- \delta}  =   \chi_{\pi- \delta}\\
\chi_{\eta- \sigma}  =   \chi_{\pi- \sigma}
\end{cases},
\;\;\;
({\rm for}\;\;
g\neq0,
\;\;m_l\neq0\;\;{\rm and}\;\; m_s=0).
\label{chiral_axial}
\end{eqnarray} 
Remarkably, the coincidence between  the strength of the chiral and axial symmetry breaking is \textcolor{black}{realized for the QCD vacuum}  
\textcolor{black}{even if the gluonic quantum correction in the $U(1)_A$ anomaly is taken into account.}
Note that 
 the \textcolor{black}{chiral $SU(2)$} symmetry cannot be seen in the susceptibility functions due to the spontaneous chiral symmetry breaking.
At the nontrivial limit in Eq.~(\ref{chiral_axial}), the quantum $U(1)_A$ anomaly contribution in the associated meson channels is disentangled from the spontaneous-chiral breaking in the meson susceptibility functions.

Once the strange quark \textcolor{black}{obtains a finite mass}, 
the topological susceptibility takes a nonzero value and gives the interference for the correlation between the chiral symmetry breaking in $\chi_{\eta- \delta}$ ($\chi_{\pi-\sigma}$) and the axial symmetry breaking in $\chi_{\pi- \delta}$ ($\chi_{\eta-\sigma}$)
through Eq.~(\ref{WI-def}). Namely, the coincidence between the chiral and axial symmetry breaking
in Eq.~(\ref{chiral_axial}) is {\it spoiled} by 
the finite strange quark mass through nonzero topological susceptibility.
This is how 
in the real-life QCD, the sizable discrepancy between the chiral and axial symmetry breaking emerges \textcolor{black}{due to a sufficiently large current mass of the strange quark controlling}
the presence of $\chi_{\rm top}$. 
Intriguingly,
given the \textcolor{black}{existence of
the nontrivial} coincidence in 
Eq.~(\ref{chiral_axial}), 
we may identify the topological susceptibility controlled by the strange quark mass
as an indicator for the discrepancy between the chiral and axial symmetry breaking strength in the meson susceptibility functions. 
\textcolor{black}{Moreover}, the nontrivial  coincidence in Eq.~(\ref{chiral_axial}) persists even at finite temperatures. This implies that the chiral symmetry is simultaneously restored with the axial symmetry 
in hot QCD at the massless limit of the strange quark.
The simultaneous symmetry restoration occurs regardless of the order of the chiral phase transition.
Thus, Eq.~(\ref{chiral_axial}) is a key limit to give us 
a new aspect of the chiral and axial phase transitions in QCD, which would help deepen our understanding of the QCD phase structure.

\begin{figure}[htbp]
\begin{tabular}{cc}
    \includegraphics[width=7.4cm]{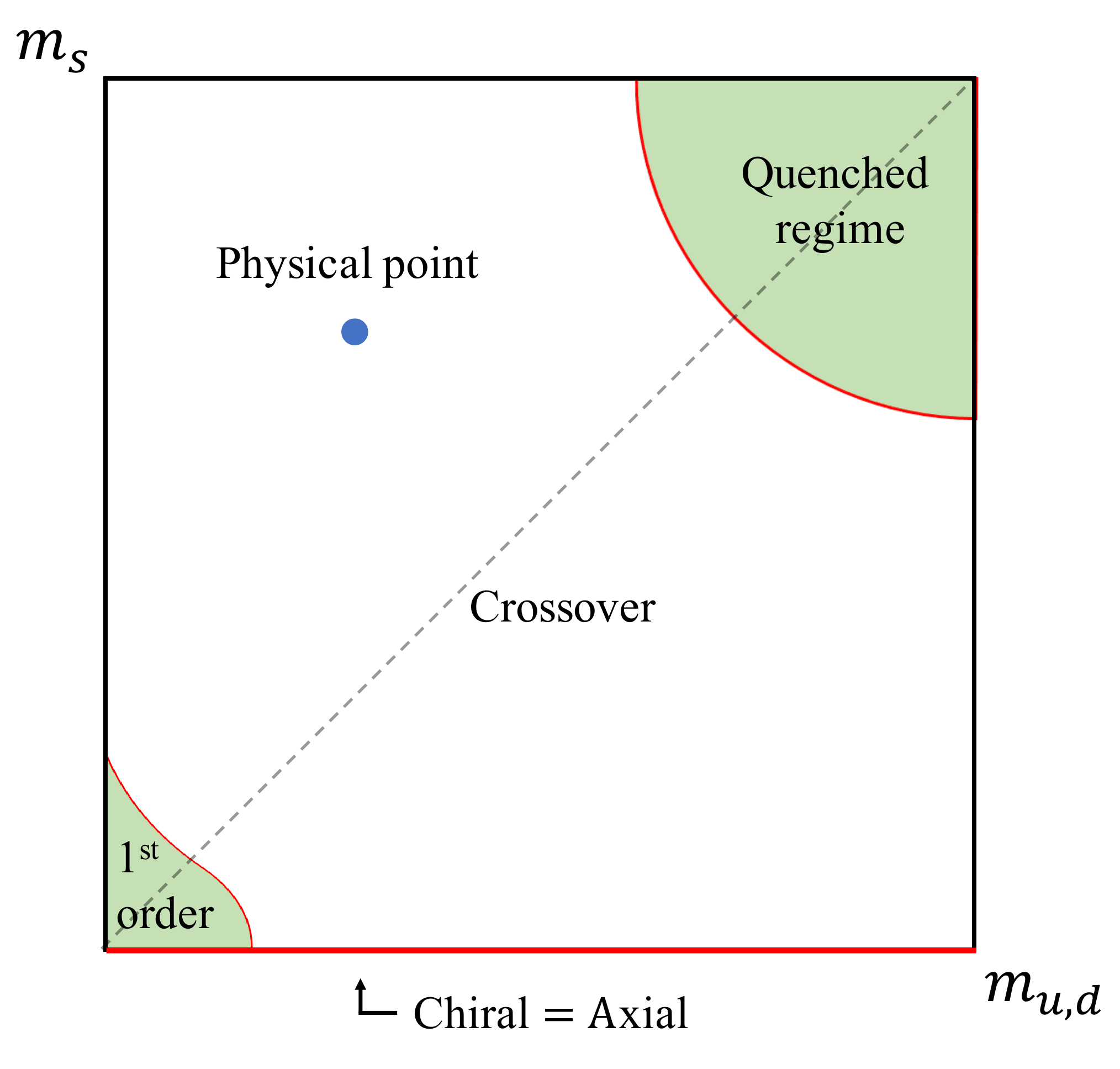}
\end{tabular}
\caption{
Conventional Columbia plot.
The strength of chiral symmetry breaking coincides with the axial symmetry breaking strength in the meson susceptibility functions with $m_s=0$ 
due to the vanishing topological susceptibility. Thus,
the simultaneous symmetry restoration between 
\textcolor{black}{the chiral $SU(2)$} and $U(1)_A$ is realized
on the $m_{u,d}$ axis
\textcolor{black}{
independently of the order of the chiral phase transition.
} 
}
\label{columbia}
\end{figure}

With those preliminaries, we may now pay attention to the Columbia plot 
to consider the quark mass dependence on the symmetry restoration.
Figure~\ref{columbia} shows the conventional Columbia plot where
the QCD phase diagram is described on 
the $m_{u,d}$-$m_s$ plane. 
As a result of Eq.~(\ref{chiral_axial}), 
the chiral symmetry is simultaneously restored with the axial symmetry on the $m_{u,d}$-axis. 

In the next section,
\textcolor{black}{in order to explain the implication} of the nontrivial coincidence in Eq.~(\ref{chiral_axial}) 
on the chiral-axial phase diagram, 
we will investigate 
the interference of the topological susceptibility for the chiral and axial symmetry breaking based on the NJL model.
\textcolor{black}{
This will be visualized by drawing the chiral-axial phase diagram in the $m_{u,d}$-$m_s$ plane, describing the trend of the chiral and axial symmetry restoration.
}

\section{
Chiral and axial symmetry breaking in low-energy QCD description
}

\subsection{
Nambu-Jona-Lasinio model}

\textcolor{black}{
To make the qualitative understanding of the nontrivial correlation among susceptibility functions in Eq.~(\ref{WI-def}) more explicit, we employ an NJL-model analysis, based on the chiral symmetry and the $U(1)_A$ symmetry
of the underlying QCD possessions.
}
As will be mentioned later, 
the NJL model predictions are shown to be in good agreement with the lattice QCD results at finite temperatures.  
In this subsection, we briefly introduce the NJL model description.
Later, we will show 
the 
formulae for the susceptibility functions in the framework of the NJL model approach.

The NJL model-Lagrangian with three quark flavors is given as
\begin{align}
	\mathcal{L}& =
	\bar{q}(i\gamma_{\mu}\partial^{\mu} -{\bm m} )q+\mathcal{L}_{4f}+\mathcal{L}_{\rm KMT} \,. 
	\label{Lag:NJL}
\end{align}
The four-quark interaction term ${\cal L}_{4f}$ is invariant under the chiral $U(3)_L\times U(3)_R$ transformation: 
$q \to U \cdot q$ where $U= \exp[ - i \gamma_5 \sum_{a=0}^8  (\lambda^a/2) \theta^a ]$ with 
$\lambda_a$ ($a=0,1,\cdots,8$)
being the Gell-Mann matrices in the \textcolor{black}{flavor space together with}  
$\lambda_0=\sqrt{2/3}\cdot 1_{3\times 3}$ \textcolor{black}{and $\theta^a$ the chiral phases}. It takes the form
\begin{align}
\mathcal{L}_{4f}&=\frac{g_s}{2}\sum^{8}_{a=0}\lbrack(\bar{q}\lambda^a q)^2+(\bar{q}i \gamma_5 \lambda^a q )^2\rbrack \,,     
\end{align}
where $g_s$ is the coupling constant.

In the NJL model approach, the anomalous $U(1)_A$ part 
is described by 
the determinant form, called  the Kobayashi-Maskawa-‘t Hooft (KMT) term \cite{Kobayashi:1970ji,Kobayashi:1971qz,tHooft:1976rip,tHooft:1976snw},
\begin{align}
\mathcal{L}_{\rm KMT}&=g_D\lbrack \mathop{\rm det}
\bar{q}(1+\gamma_5 )q+ {\rm h.c.} \rbrack 
\,
\end{align}
with 
the \textcolor{black}{constant real parameter} $g_D$. 
Note that the axial breaking $\mathcal{L}_{\rm KMT}$, which would be induced from the QCD instanton configuration, 
still keeps the chiral $SU(3)_L\times SU(3)_R$ symmetry.

	
The current conservation of 
the chiral symmetry and the $U(1)_A$ symmetry becomes anomalous due to 
the presence of the quark mass terms and the KMT term: 
\begin{eqnarray}
\partial_\mu j_A^{a\mu }&=&
i\bar q\left\{\bm m, \frac{\tau^a}{2}   \right\}\gamma_5 q,
\nonumber\\
\partial_\mu j^\mu_A&=&2i \bar q {\bm m}\gamma_5 q
-4N_fg_D {\rm Im} \left[{\rm det}\bar q(1-\gamma_5)q  \right],
\label{current_cons_law_NJL}
\end{eqnarray}
\textcolor{black}{where the curly brackets $\{,\}$ represents an anticommutator.}
In the spirit of effective models based on the underlying QCD,
the anomalous conservation laws of the NJL model have to be linked \textcolor{black}{with those of the underlying QCD}, in Eq.~(\ref{current_cons_law_QCD}).
Thus,
the KMT operator, $g_D {\rm Im} \left[{\rm det}\bar q(1-\gamma_5)q  \right]$, may mimic the $U(1)_A$ anomaly of the gluonic operator, $g^2 \epsilon^{\mu\nu\rho\sigma}  F _{\mu\nu}^a  F^a_{\rho\sigma}$.  
One \textcolor{black}{should notice here} that 
the $U(1)_A$ anomaly described by the KMT term can vanish by taking $g_D \to 0$. 
As far as the $U(1)_A$ anomaly contribution is concerned, 
this limit corresponds to turning off the QCD 
gauge coupling $g$, \textcolor{black}{
which is equivalent to the trivial
} 
limit for the vanishing axial anomaly~\footnote{ 
Note that even in the case of the vanishing anomaly associated with $g_D=0$ the NJL model is still an interacting theory due to the existence of the four-quark interaction term.
Although $g_D$ is not completely compatible with the QCD coupling constant $g$, 
we can monitor the $U(1)_A$ anomaly contribution
through the effective coupling constant $g_D$.
}.


\subsection{Mean-field approximation and vacuum of NJL model}
In this work, we employ the mean-field approximation corresponding to the large $N_c$ expansion.
\textcolor{black}{
Within the mean-field approximation, 
the interaction terms go like}
\begin{align}
\mathcal{L}_{4f}&
\to 
2 g_s\left(\alpha\bar uu+
 \beta \bar dd+
 \gamma \bar ss
 \right)
 - g_s(\alpha^2+\beta^2 +\gamma^2)
\,,     \nonumber\\
\mathcal{L}_{\rm KMT}
&\to
2 g_D \left(
\beta\gamma\bar uu
+\alpha\gamma\bar dd
+\alpha\beta \bar ss
\right)
-4 g_D\alpha\beta\gamma,
\end{align}
\textcolor{black}{where $\alpha,\beta$, and $\gamma$ denote the quark condensates,}
	 \begin{equation}
	 	\langle \bar u u \rangle \equiv \alpha, 
	 	\quad \langle \bar d d \rangle \equiv \beta, 
	 	\quad \langle \bar s s \rangle \equiv \gamma, 
	 \end{equation} 
In the isospin symmetric limit ($m_u=m_d=m_l$),
$\alpha$ and $\beta$ are taken as
$\alpha=\beta$.	 
\textcolor{black}{
Hence, the NJL Lagrangian is reduced to the mean-field Lagrangian ${\cal L}_{\rm mean}$: 
}
\begin{eqnarray}
&&{\cal L}=
\bar q(i\gamma^\mu \partial_\mu -{\bm m})q  
+
\frac{g_s}{2}
\left[
(\bar q\lambda_a q)^2
+
(\bar q i\gamma_5\lambda_a q)^2
\right]
+
g_D{\rm det} [   \bar q (1+\gamma_5) q +{\rm h.c.}] 
\nonumber\\
&\to&
{\cal L}_{\rm mean}=
\bar q(i\gamma^\mu \partial_\mu -{\bm M})q  
 -g_s(\alpha^2+\beta^2 +\gamma^2)
 -4g_D\alpha\beta\gamma,
\end{eqnarray}
\textcolor{black}{
where ${\bm M} = {\rm diag}(M_u,M_d,M_s)$ represents the mass matrix of the dynamical quarks,}
	\begin{align} 
		M_u&=m_u-2g_s \alpha - 2g_D\beta \gamma \,, \notag\\
		M_d&=m_d-2g_s \beta - 2g_D\alpha \gamma \, , \notag\\
		M_s&=m_s-2g_s \gamma - 2g_D\alpha \beta \,. 
	\end{align}

By integrating out the quark field in the generating functional of the mean-field Lagrangian, 
\textcolor{black}{the effective potential at finite temperature 
is evaluated as {(see e.g.,~\cite{Klevansky:1992qe})}
}
\begin{eqnarray}
V_{\rm eff}(\alpha,\beta,\gamma) =
g_s(\alpha^2+\beta^2 +\gamma^2)
 +4g_D\alpha\beta\gamma
-2N_c\sum_{i}
\int^\Lambda \frac{d^3p}{(2\pi)^3}
\left\{
E_i
+2T\ln\left(
1+e^{- E_i/T      }
\right)
\right\},
\label{effpot}
\end{eqnarray}
where \textcolor{black}{
$N_c=3$ denotes the number of colors, and 
$E_i=\sqrt{M_i^2+p^2}$ are the energies of the constituent quarks. 
The NJL model is a nonrenormalizable theory. Hence the momentum cutoff $\Lambda$ 
should be prescribed 
in the quark loop calculation 
to regularize the ultraviolet divergence.
In Eq.(\ref{effpot}) 
we have applied a sharp cutoff regularization to the three-dimensional momentum integration.  
}

\textcolor{black}{
The quark condensates sit on the stationary point of the effective potential with respect to $\alpha$, $\beta$, and $\gamma$, which are determined from the stationary conditions for the
effective potential,
}
\begin{eqnarray}
\frac{\partial V_{\rm eff}(\alpha,\beta,\gamma)}{\partial \alpha}=0,\;\;\;
\frac{\partial V_{\rm eff}(\alpha,\beta,\gamma)}{\partial \beta}=0
,\;\;\;
\frac{\partial V_{\rm eff}(\alpha,\beta,\gamma)}{\partial \gamma}=0.
\end{eqnarray}
\textcolor{black}{
By solving the stationary conditions, one can obtain the following analytic expression of the quark condensates, which corresponds to the quark one-loop result:
}
\begin{equation}
		\langle  \bar q_i q_i \rangle =- 2N_c\int^{\Lambda}\frac{d^3 p }{(2\pi)^3}  \frac{M_i}{E_{i}}\big{[}1-2 (\exp (E_{i}/T)+1)^{-1}\big{]}.  
		\label{qcond}
\end{equation}

\subsection{Scalar- and pseudoscalar-meson susceptibility in NJL model}
In this subsection, we introduce the pseudoscalar meson susceptibilities in the NJL model approach.
First, the pseudoscalar susceptibilities,
which construct the meson susceptibilities in Eq.(\ref{chi-meson--def}), are evaluated as \cite{Hatsuda:1994pi}
\begin{eqnarray}
\chi_P^{ab}(\omega =0, {\bm p} =0)
 &=& \lim_{p\to 0} 
 \int_T d^4 x e^{ip\cdot x}
 \langle (i \bar q (x) \gamma_5 \lambda^a  q (x))(i \bar q (0)\gamma_5 \lambda^b  q(0)) \rangle
 \,,
\end{eqnarray}
with the external momentum $p^\mu = (\omega,{\bm p})$. 
\textcolor{black}{
The susceptibilities are defined by the two-point correlation function of the quark bilinear field at the zero external momentum, $p^\mu=0$.
}

\textcolor{black}{
We work in the Random Phase Approximation~\cite{Hatsuda:1994pi}, so that the pseudoscalar susceptibilities are evaluated only through the resummed polarization diagram of the quark loop, taking into account the four-point interactions in the NJL model.
Within the mean-field approximation, these four-point interaction terms represent fluctuations from the vacuum characterized by the nonzero quark condensates, 
\begin{eqnarray}
{\cal L}_{{\rm fluc}}^{(4)}
&=&
\frac{1}{2}\left(g_s +\frac{2}{3}(\alpha+\beta +\gamma)g_D    \right)   
\bar q    q  \bar q   q
+
\frac{1}{2}(g_s - g_D \gamma)  
\sum_{i=1}^3
\bar q  \lambda^i  q  \bar q  \lambda^i  q
+
\frac{1}{2}\left(g_s +\frac{1}{3}(\gamma -2\alpha -2\beta )g_D \right)  
\bar q  \lambda^8  q  \bar q  \lambda^8  q
\nonumber\\
&&
+\left( 
\frac{\sqrt{2}}{6} (2\gamma - \alpha - \beta ) g_D
\right)
\bar q    q  \bar q  \lambda^8  q
+ \left(
-\frac{g_D}{\sqrt{6}}(\alpha- \beta)
\right)  \bar q    q  \bar q  \lambda_3 q
+
\left(
\frac{g_D}{\sqrt{3}}(\alpha- \beta)
\right) 
\bar q   \lambda_3  q  \bar q \lambda_8  q\nonumber\\
&&
+
\frac{1}{2}\left(g_s -\frac{2}{3}(\alpha+\beta +\gamma) g_D   \right)  
\bar q i\gamma_5   q  \bar q i\gamma_5  q
+
\frac{1}{2}(g_s + g_D \gamma)  
\sum_{i=1}^3
\bar q i\gamma_5 \lambda^i  q  \bar q i\gamma_5 \lambda^i  q
\nonumber\\
&&
+
\frac{1}{2}\left(g_s -\frac{1}{3}(\gamma -2\alpha -2\beta )g_D \right)
\bar q i\gamma_5 \lambda^8  q  \bar q i\gamma_5 \lambda^8  q
\nonumber\\
&&
+\left( 
-\frac{\sqrt{2}}{6} (2\gamma - \alpha - \beta ) g_D
\right)
\bar q  i\gamma_5  q  \bar q i\gamma_5 \lambda^8  q
+ 
\left(
\frac{g_D}{\sqrt{6}}(\alpha- \beta)
\right) 
\bar q   i\gamma_5 q  \bar q i\gamma_5 \lambda_3 q
+
\left(
\frac{g_D}{\sqrt{3}}(\alpha- \beta)
\right) 
\bar q  i\gamma_5    \lambda_3 q \bar q i\gamma_5 \lambda_8  q\nonumber\\
&&
+\cdots.
\label{fluc_four_int}
\end{eqnarray}
Here we have picked up only the interaction terms relevant to $\chi_{\pi,\eta,\delta,\sigma}$. 
Note that since we keep the isospin symmetry, i.e., $\alpha=\beta$, the four-point interaction terms proportional to $(\alpha-\beta)$ vanish, hence those terms will not come into play in the later discussion. 
}
\textcolor{black}{
Then the pseudoscalar susceptibilities are expressed as
}
\begin{eqnarray}
\chi_{P}^{ac} &=&- 
\lim_{\omega\to0}
 \lim_{{\bm p}\to 0}
\Pi_{P}^{ab }(\omega,{\bm p}) D^{-1}_{P bc} (\omega,{\bm p}),
\end{eqnarray}
with 
\begin{eqnarray}
D^{ab}_{P}(\omega,{\bm p})&=& \delta^{ab} + G_{P}^{ac}\Pi_{P}^{cb }(\omega,{\bm p}),
\end{eqnarray}
where $G_P^{ab}$ is the coupling strength corresponding to the four-point interaction within the mean field approximation \textcolor{black}{and  $\Pi_P^{ab}$ is the polarization function at the quark one-loop level. 
Note that $\chi_{P}^{ab}$ $G_{P}^{ab}$ and $\Pi_{P}^{ab }$ take the matrix form. 
}

\textcolor{black}{
The pion susceptibility corresponds to this $\chi_P^{ab}$ with  $a,b=1,2,3$ as 
\begin{eqnarray}
\chi_\pi 
&=&\chi_{P}^{11} 
\nonumber\\
&=&
- \left[ {\bf  \Pi_{\pi}} (0,0) \cdot \Bigl\{ {\bf 1}_{3\times3} +{\bf G_{\pi}}\cdot {\bf \Pi_{\pi}(0,0) }\Bigl\}^{-1} \right]^{11}
=\chi_{P}^{22}=\chi_{P}^{33}
\,, 
	\label{pion}
\end{eqnarray}
where 
the coupling strength in the pion channel ${\bf G}_\pi={\rm diag}(G_P^{11},G_P^{22},G_P^{33})$
and the pion polarization function ${\bf \Pi}_\pi ={\rm diag}(\Pi_P^{11},\Pi_P^{22},\Pi_P^{33})$ are given as
\begin{eqnarray}
{\bf G_{\pi}} &=& (g_s + g_D \gamma) {\bf 1}_{3\times3},
\nonumber\\
{\bf \Pi_{\pi} }&=&(I_P^{u}+I_P^{d} ){\bf 1}_{3\times3}=2I_P^{u} {\bf 1}_{3\times3}
\,, 
\end{eqnarray}
with $I_P^{i}(\omega, \boldsymbol{p})$ being the pesudoscalar one-loop polarization functions~\cite{Kunihiro:1991hp}, 
\begin{equation}
	I_P^{i}(0,0)=-\frac{N_c}{\pi ^2}\int^\Lambda _0  d p\, p^2  \frac{1}{E_{i}}\left[1 -2 \left( \exp(E_i /T)+1\right)^{-1} \right]\,, \qquad {\rm for } \quad i=u,d,s 
	\,. \label{IPij}
\end{equation} 
Note that owing to the isospin symmetry, ${\bf G}_{\pi}$ and ${\bf \Pi}_{\pi}$ exhibit no off-diagonal components.   
In contrast, as shown in Eq.~(\ref{fluc_four_int}),the flavor symmetry breaking associated with the $U(1)_A$ anomaly provides off-diagonal components in $G_{P}^{ab}$ and $\Pi_{P}^{ab }$ for $a,b=0,8$,
\begin{equation}
{\bf G_P}=
\begin{pmatrix}
	G_P^{00} & G_P^{08}\\
	G_P^{80} & G_P^{88}
\end{pmatrix}
=
\begin{pmatrix}
	g_s-\frac{2}{3}(\alpha +\beta+\gamma )g_D &  -\frac{\sqrt{2}}{6} (2\gamma-\alpha-\beta )g_D \\
	-\frac{\sqrt{2}}{6} (2\gamma-\alpha-\beta )g_D & g_s-\frac{1}{3}(\gamma-2\alpha -2\beta )g_D
\end{pmatrix}
\,, \label{Gs-p}
\end{equation}
\begin{equation}
{\bf \Pi_P}=
\begin{pmatrix}
	\Pi _P^{00} & \Pi _P^{08}\\
	\Pi _P^{80} & \Pi _P^{88}
\end{pmatrix}
=
\begin{pmatrix}
	\frac{2}{3}	(2I_P^{u}+I_P^{s}) &  \frac{2\sqrt{2}}{3}(I_P^{u}-I_P^{s}) \\\frac{2\sqrt{2}}{3}(I_P^{u}-I_P^{s} ) & \frac{2}{3} (I_P^{u} + 2I_P^{s})
\end{pmatrix}
\,. 
\end{equation}
The pseudoscalar susceptibilities in Eq.(\ref{f_chi}) are obtained as linear combinations of $\chi_P^{ab}$ for $a,b=0,8$ as  
}
\begin{equation}
	\begin{pmatrix}
	\frac{1}{2}\chi_P^{uu}
	+\frac{1}{2}\chi_P^{ud}
	\\
	\chi_P^{us} \\
	\chi_P^{ss}
	\end{pmatrix}
=
	\begin{pmatrix}
	\frac{1}{6} & \frac{\sqrt2}{6} & \frac{1}{12} \\
    \frac{1}{6} & -\frac{\sqrt2}{12} & -\frac{1}{6} \\
    \frac{1}{6} & -\frac{\sqrt2}{3} & \frac{1}{3}
	\end{pmatrix}
	\begin{pmatrix}
	\chi_P^{00} \\
	\chi_P^{08} \\
	\chi_P^{88}
	\end{pmatrix}
	\label{chip}
	\,, 
\end{equation}
where we have taken the isospin symmetric limit into account, i.e., $\chi_P^{uu} = \chi_P^{dd}$ and $\chi_P^{us}=\chi_P^{ds}$. 
Then the $\eta$ meson susceptibility is evaluated as 
\begin{equation} 
	\chi_{\eta}= 
	2 \chi_P^{uu}+2 \chi_P^{ud}
	\label{eta}
	\,. 
\end{equation}


Similarly 
the scalar meson susceptibilities are given by 
\begin{eqnarray}
\chi_{S}^{ac}
&=&
-\Pi_{S}^{ab}(0,0)
D_{Sbc}^{-1} (0,0)
\,, 
\end{eqnarray}
with
\begin{eqnarray}
D_S^{ab} (0,0) =  \delta^{ab} +  G_{S}^{ac}\Pi^{cb}_{S}(0,0)
\end{eqnarray}
where $G_{S}^{ab}$ is the coupling strength matrix and $\Pi_{S}^{ab}$ is the polarization tensor matrix in the scalar channel. 

\textcolor{black}{
The explicit formula for $\chi_\delta$ reads 
\begin{eqnarray} 
\chi_{\delta}&=&\chi_{S}^{11}\nonumber\\
	&=&\left[-{\bf \Pi}_{\delta}(0,0)\cdot 
         \left\{ {\bf 1}_{3\times 3}+{\bf G}_{\delta}\cdot{\bf \Pi}_{\delta}(0,0)\right\}^{-1}\right]^{11}=
         \chi_{S}^{22}
         =\chi_{S}^{33}
	\label{delta}
	\,, 
\end{eqnarray}
where 
the coupling strength in the $\delta$ meson channel ${\bf G}_\delta ={\rm diag}(G_S^{11}, G_S^{22}, G_S^{33})$
and the $\delta$-meson polarization function ${\bf \Pi}_\delta  ={\rm diag}(\Pi_S^{11}, \Pi_S^{22}, \Pi_S^{33})$ are given as
\begin{eqnarray}
{\bf G}_{\delta} &=& (g_s - g_D\gamma ) {\bf 1}_{3\times 3},\nonumber\\
{\bf \Pi}_{\delta}&=&(I_S^{u}+I_S^{d}){\bf 1}_{3\times 3}=2I_S^{u} {\bf 1}_{3\times 3},
\end{eqnarray}
with
$I_S^{i}$ being the scalar one-loop polarization functions, 
\begin{equation}
	I_S^{i}(0,0)=-\frac{N_c}{\pi ^2}\int^\Lambda _0 p^2 dp \frac{E_{i}^2-M_i^2}{E_i^3}\lbrace1-2[\exp(E_i /T)+1]^{-1}\rbrace \,, \qquad i=u,d,s
\,. 
\end{equation} 
For $a,b=8$, $G_S^{ab}$ and $\Pi_S^{ab}$ are given as
\begin{equation}
	{\bf G}_S=
\begin{pmatrix}
	G_S^{00} & G_S^{08}\\
	G_S^{80} & G_S^{88}
\end{pmatrix}
=
\begin{pmatrix}
	g_s+\frac{2}{3}(\alpha +\beta+\gamma )g_D &  \frac{\sqrt{2}}{6} (2\gamma-\alpha-\beta )g_D \\
	\frac{\sqrt{2}}{6} (2\gamma-\alpha-\beta )g_D & g_s+\frac{1}{3}(\gamma-2\alpha -2\beta )g_D
\,, 
\end{pmatrix}
\end{equation}
\begin{equation}
{\bf \Pi_S}=
\begin{pmatrix}
	\Pi _S^{00} & \Pi _S^{08}\\
	\Pi _S^{80} & \Pi _S^{88}
\end{pmatrix}
=
\begin{pmatrix}
	\frac{2}{3}	(2I_S^{u}+I_S^{s}) &  \frac{2\sqrt{2}}{3}(I_S^{u}-I_S^{s}) \\\frac{2\sqrt{2}}{3}(I_S^{u}-I_S^{s} ) & \frac{2}{3} (I_S^{u} + 2I_S^{s})
\,.   
\end{pmatrix}
\end{equation}
Then, taking the linear combinations of $\chi_S^{ab}$,
one can obtain the scalar susceptibilities in Eq.~(\ref{f_chi_S}),
}
\begin{equation}
	\begin{pmatrix}
	\frac{1}{2}\chi_S^{uu}
	+\frac{1}{2}\chi_S^{ud}
	 \\
	\chi_S^{us} \\
	\chi_S^{ss}
	\end{pmatrix}
=
	\begin{pmatrix}
	\frac{1}{6} & \frac{\sqrt2}{6} & \frac{1}{12} \\
    \frac{1}{6} & -\frac{\sqrt2}{12} & -\frac{1}{6} \\
    \frac{1}{6} & -\frac{\sqrt2}{3} & \frac{1}{3}
	\end{pmatrix}
	\begin{pmatrix}
	\chi_S^{00} \\
	\chi_S^{08} \\
	\chi_S^{88}
	\end{pmatrix},
\label{scalar:mix:chi}
\end{equation}
where the isospin symmetric limit has taken into account,
$\chi_S^{uu}=\chi_S^{dd}$
and $\chi_S^{us}=\chi_S^{ds}$. 
From Eq.(\ref{scalar:mix:chi}) we obtain the sigma-meson susceptibility $\chi_{\sigma}$ in the NJL model analysis,
\begin{equation}
	\chi_{\sigma}= 
	2 \chi_S^{uu}+2 \chi_S^{ud}.
	\label{sigma}
\end{equation}

\subsection{
Trivial and nontrivial coincidence of chiral and axial breaking 
in a view of the NJL description
}

With 
Eq.~\ref{chitop_wfac} and 
the pseudscalar meson susceptibility in Eqs.~(\ref{pion}) and (\ref{eta}),
the topological susceptibility in the NJL model can be described as
\begin{eqnarray}
\chi_{\rm top}=-m_lm_s g_D 
\left(
\alpha\frac{  (\Pi_P^{08})^2-\Pi_P^{00} \Pi_P^{88}  }{6{\rm det }(1+G_P \Pi_P)}
\right).
\label{chitop_mlmsgd}
\end{eqnarray}
Using Eq.~(\ref{chitop_mlmsgd}) together with 
the meson susceptibilities in Eqs.~(\ref{pion}), (\ref{eta}), (\ref{delta}) and (\ref{sigma}), one can easily 
check that the NJL model reproduces Eq.~(\ref{WI-def}): 
\begin{eqnarray*}
&&\frac{1}{4}m_l^2 (\chi_{\eta-\delta}-\chi_{\pi-\delta})=
\chi_{\rm top},\nonumber\\
&&\frac{1}{4}m_l^2 (\chi_{\eta-\sigma}-\chi_{\pi-\sigma})=
\chi_{\rm top}.
\end{eqnarray*}
Note that the analytical expression of $\chi_{\rm top}$ in Eq.~(\ref{chitop_mlmsgd}) explicitly shows that 
$\chi_{\rm top}$ is proportional to the $U(1)_A$ 
anomaly-related coupling 
$g_D$.  
As was noted, 
$\chi_{\rm top}$ goes away
in the limit of the vanishing $U(1)_A$ anomaly, $g_D \to 0$,  while the meson susceptibility functions keep finite values.
This is the NJL-model realization of the trivial coincidence between the indicators for the chiral and axial symmetry breaking in the meson susceptibility functions, as 
in Eq.~(\ref{trivial_limit_QCD}):  
\begin{eqnarray}
\begin{cases}
\chi_{\eta- \delta}  =   \chi_{\pi- \delta}\\
\chi_{\eta- \sigma}  =   \chi_{\pi- \sigma}
\end{cases}
, \;\;\;
({\rm for}\;\; g_D=0,\;\;m_l\neq0\;\;
{\rm and}\;\; m_s\neq 0).
\label{chiral_axial_NJL}
\end{eqnarray}

Of crucial is to note also that $\chi_{\rm top}$ in Eq.~(\ref{chitop_mlmsgd})
is proportional also to both the light quark mass and the strange quark mass, 
as the consequence of the flavor singlet nature in Eq.~(\ref{FS_condition}). 
Hence, 
the NJL model also provides the nontrivial  coincidence 
between the chiral and axial indicators 
with keeping the $U(1)_A$ anomaly, as derived from the underlying QCD in Eq.~(\ref{chiral_axial}): 
\begin{eqnarray}
\begin{cases}
\chi_{\eta- \delta}  =   \chi_{\pi- \delta}\\
\chi_{\eta- \sigma}  =   \chi_{\pi- \sigma}
\end{cases}
, \;\;\;
({\rm for}\;\; g_D\neq0,\;\;m_l\neq0\;\;
{\rm and}\;\; m_s=0).
\label{chiral_axial_NJL}
\end{eqnarray} 
This coincidence 
implies that $U(1)_A$ anomaly contribution in the associated meson channels becomes 
invisible in the meson susceptibility functions at $m_s=0$ where $\chi_{\rm top}=0$, even in the presence of the $U(1)_A$ anomaly ($g_D\neq0$).


The topological susceptibility has also been studied by some effective model approaches \cite{Fukushima:2001hr,Jiang:2012wm,Jiang:2015xqz,Lu:2018ukl,GomezNicola:2019myi, DiVecchia:1980yfw,Hansen:1990yg,Leutwyler:1992yt,Bernard:2012ci,Guo:2015oxa}
However, the previous studies have not taken account of the flavor singlet nature of the topological susceptibility, so that the nontrivial  coincidence in Eq.(\ref{chiral_axial_NJL}) 
has never been addressed
\footnote{
One may further rotate quark fields by the $U(1)_A$ transformation with the rotation angles $\theta_{f=u,d,s}$, so that the NJL Lagrangian is shifted as 
${\cal L}_{\rm NJL}(\theta) \to  {\cal L}_{\rm NJL}(\theta-\bar \theta)+\bar \theta Q_{\rm NJL}
$ where $\bar \theta=\theta_u+\theta_d+\theta_s$ and 
$Q_{\rm NJL} =  - 4 g_D {\rm Im} \left[{\rm det}\bar q_i(1-\gamma_5)q_j  \right]$.
By taking the $\bar \theta =\theta$, the $\theta$-dependence is completely rotated away from the quark mass term, and then moves to the $Q_{\rm NJL}$ term:
${\cal L}_{\rm NJL}(\theta=0)+ \theta Q_{\rm NJL}$. 
Indeed, the topological susceptibility has been evaluated based on the NJL Lagrangian including the $\theta Q_{\rm NJL}$ term:
$\chi_{\rm top}=\int_T d^4x \langle Q_{\rm NJL}(x)Q_{\rm NJL}(0)\rangle$ \cite{Fukushima:2001hr}.
However,  
the $\theta$-dependence on the $\theta Q_{\rm NJL}$ term
accidentally goes away within   
the ordinary 
mean-field approach. This is
because  
$Q_{\rm NJL}$ vanishes under the mean-field approximation, $Q_{\rm NJL} \propto {\rm det}[\bar{q} (1+\gamma_5) q] - {\rm det}[\bar{q} (1- \gamma_5)q] \to 
(\beta\gamma\bar uu
+\alpha\gamma\bar dd
+\alpha\beta \bar ss
-2\alpha\beta\gamma)- (\beta\gamma\bar uu
+\alpha\gamma\bar dd
+\alpha\beta \bar ss
-2\alpha\beta\gamma)=0
$. Hence,
the topological susceptibility can not be evaluated based on ${\cal L}_{\rm NJL}(\theta=0)+ \theta Q_{\rm NJL}$ within the mean-field approach through the second derivative of the generating functional with respect to $\theta$. 
Thereby, we do not take this way, instead, directly apply the 
Eq.~(\ref{chitop_wfsc}) which is evaluated from the $\theta$-dependent quark-mass term with the flavor singlet nature.
}.

\section{
Quark mass dependence on QCD vacuum structure
}
In this section, 
through Eq.~(\ref{WI-def}), we numerically explore the correlations among  
the susceptibility functions for the chiral symmetry breaking, the axial symmetry breaking and the topological charge.

\subsection{
QCD vacuum structure with physical quark masses
}

To exhibit the numerical results of the susceptibility functions, we take the value of the parameters as listed in Table \ref{Parameter} \cite{Hatsuda:1994pi}.
With the input values, the following four hadronic observables are obtained at $T=0$ \cite{Hatsuda:1994pi}, 
\begin{equation}
	m_{\pi}=136 \, {\rm MeV}, \quad f_{\pi}=93 \, {\rm MeV}, 
	\quad m_K=495.7 \, {\rm MeV}, \quad m_{\eta \prime}=957.5\, {\rm MeV} 
\,, 
\end{equation}
which are in good agreement with the experimental values.
Furthermore, the topological susceptibility at the vacuum ($T=0$) qualitatively agrees with the lattice observations \cite{Borsanyi:2016ksw,Bonati:2018blm}, as discussed in \cite{Cui:2021bqf}. 

\begin{table}[!htbp]
\centering
\caption{Parameter setting}
\begin{tabular}{|c|c|}
\hline
\textbf{parameters} & \textbf{values}               \\ \hline
light quark mass $m_l$   & 5.5 MeV                       \\ \hline
strange quark mass $m_s$  & 138 MeV                       \\ \hline
four-fermion coupling constant $g_s$                & 0.358 ${\rm fm}^2$   \\ \hline
six-fermion coupling constant $g_D$                & $-$ 0.0275 ${\rm fm}^5$ \\ \hline
cutoff $\Lambda$             & 631.4 MeV                     \\ \hline
\end{tabular}
\label{Parameter}
\end{table}

We will not consider intrinsic-temperature dependent couplings, 
instead, all the $T$ dependence should be induced only 
from the thermal quark loop corrections.
Actually, the present NJL shows good 
agreement with lattice QCD results on the temperature scaling 
for the chiral, axial, and topological susceptibilities, 
as shown in Ref.~\cite{Cui:2021bqf}. 
In this sense, we do not need to introduce such an 
intrinsic $T$ dependence for the model parameters 
in the regime up to temperatures around the 
chiral crossover.

\begin{figure}[htbp]
\begin{tabular}{cc}
   \begin{minipage}{0.5\hsize}
     \begin{center}
      \includegraphics[width=9.1cm]{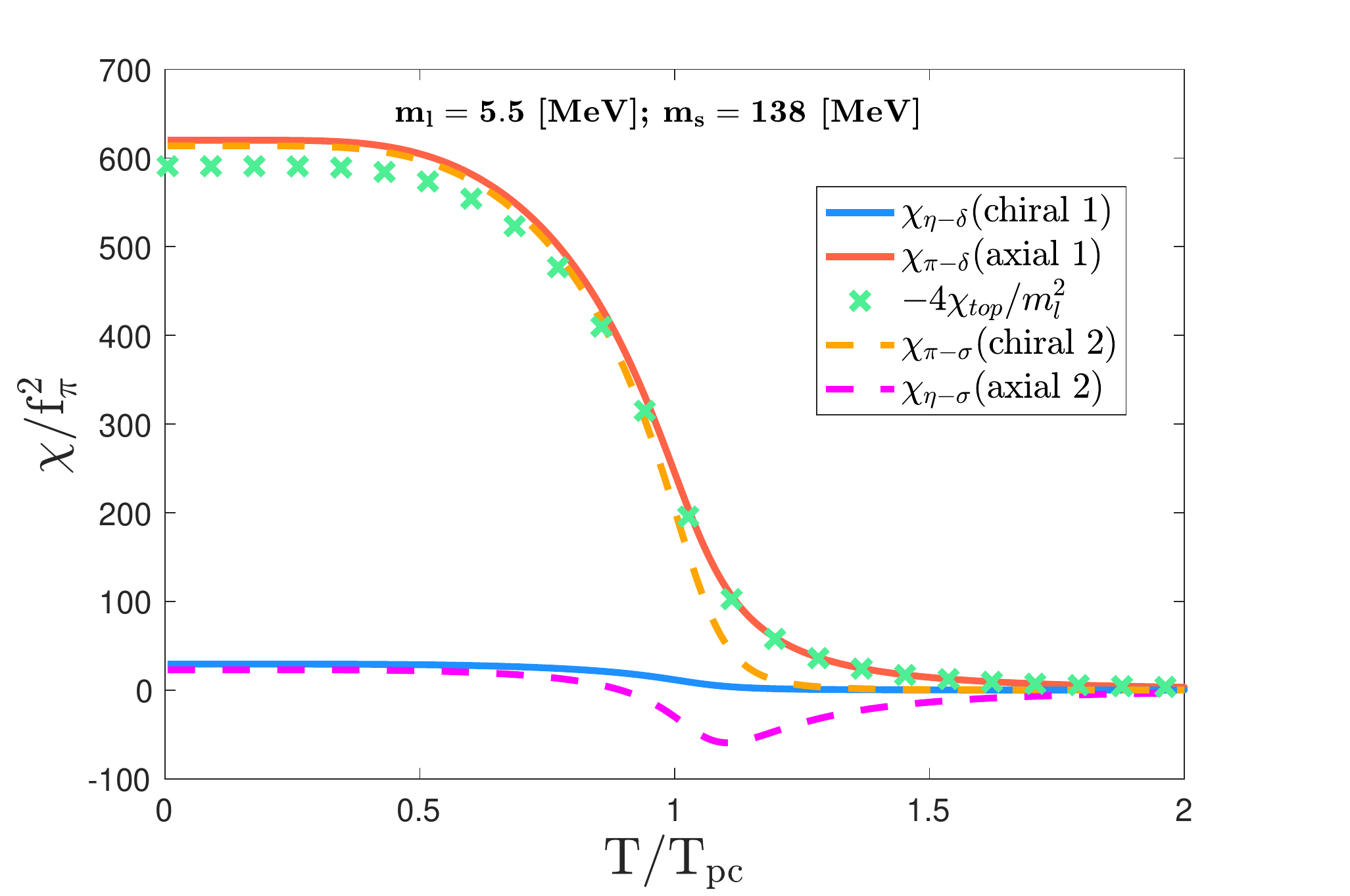}
    \subfigure{(a)}
      \end{center}
    \end{minipage}
    \begin{minipage}{0.5\hsize}
     \begin{center}
     \includegraphics[width=8.44cm]{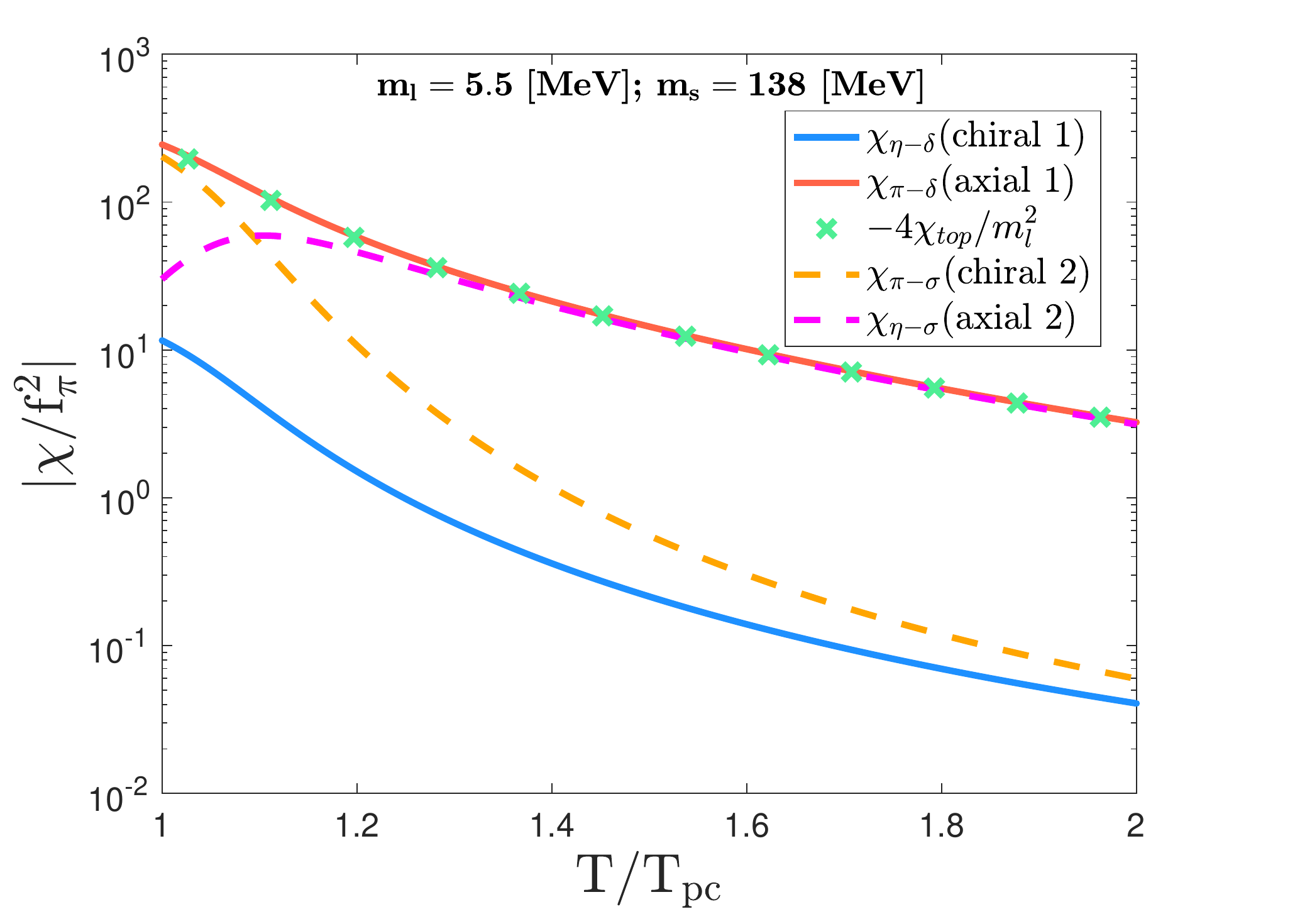}
     \subfigure{(b)}
    \end{center}
\end{minipage}
\end{tabular}
\caption{
The temperature dependence of the susceptibility functions at the physical point  
($m_l=5.5$ MeV and $m_s=138$ MeV) for (a)
$T/T_{\rm pc}=0-2$ 
and
(b) $T/T_{\rm pc}=1-2$.
The pseusocritical temperature for the 
chiral crossover has been observed to be 
$T_{\rm pc}\simeq 189$ MeV.
The susceptibility functions have been normalized by  
square of the pion decay constant ($\simeq 93$ MeV), and 
the temperature axis also by $T_{\rm pc}$, so that 
all quantities are dimensionless to reduce the systematic uncertainty (approximately about 30\%) associated with the present NJL model description of QCD. See also the text.
}
\label{crossover_physical}
\end{figure}

In Fig.~\ref{crossover_physical}, we first show plots of the susceptibilities as a function of temperature.
This figure shows that the meson susceptibilities forming the chiral partners (chiral indicators),  
$\chi_{\eta-\delta}$ and $\chi_{\pi-\sigma}$,
smoothly approach zero at high temperatures, but do not exactly reach zero. This tendency implies that the NJL model undergoes a chiral crossover~\footnote{What we work on are the susceptibilities, which correspond to meson-correlation functions 
at zero momentum transfer. 
This is in contrast to the conventional meson correlators depending on the transfer momentum, 
from which meson masses are read off.
Furthermore, 
the susceptibilities involve contact term contributions independent of momenta, which could be sensitive to a high-energy scale physics, while the conventional meson correlators are dominated by the low-lying meson mass scale. 
Nevertheless, the degeneracy of the chiral or axial partners at high temperatures, similar to those detected in the susceptibility, can also be seen in the mass difference or equivalently the degeneracy of the conventional meson correlators for the partners, which is simply because the mass difference plays an alternative indicator of the chiral  or axial breaking 
as observed in the lattice simulation~\cite{Brandt:2016daq}.  }.

The pseudocritical temperature of the chiral crossover can be evaluated  from the inflection point of $\chi_{\eta-\delta}$ or $\chi_{\pi-\sigma}$ with respect to temperature, 
\textcolor{black}{$\left.\mathrm{d}^2 \chi_{\eta-\delta,\pi-\sigma}/\mathrm{d}T^2\right|_{T=T_{\mathrm{pc}}}=0$},
and then we find $T_{\rm pc}|_{\rm NJL}\simeq 189$ MeV. 
This inflection point coincides with that estimated from the light quark condensate \cite{Cui:2021bqf}.
However the NJL's estimate of the pseudocritical temperature is somewhat bigger than the lattice QCD's, $T_{\rm pc}|_{\rm lat.}\sim 155$ MeV \cite{Aoki:2009sc,Bhattacharya:2014ara, Ding:2015ona}. 
In fact, the NJL analysis is implemented in 
the mean field approximation corresponding to the large $N_c$ limit. 
The corrections of the beyond mean field approximation would be subject to the size of 
the next-to-leading order corrections of the large $N_c$ expansion, $O(1/N_c)\sim O(0.3)$.  
Including the possible corrections to the current model analysis, the NJL's result might be consistent with the lattice observation. 
Supposing this systematic deviation by about $30 \%$ to be accepted within the framework of the the large $N_c$ expansion, one may say that the NJL description at finite temperatures yields qualitatively good agreement with the lattice QCD simulations. 
Indeed, 
all the temperature dependence of $\chi_{\eta-\delta}$, $\chi_{\pi-\delta}$, and $\chi_{\rm top}$ qualitatively accords with the lattice data \cite{Buchoff:2013nra,Bhattacharya:2014ara,Borsanyi:2016ksw,Bonati:2018blm,Petreczky:2016vrs} 
(for the detailed discussion, see \cite{Cui:2021bqf}).



From the panel (a) of Fig.~\ref{crossover_physical},
one can see a sizable difference in the meson susceptibilities
between 
the chiral indicator $\chi_{\eta-\delta}$ 
and the axial indicator $\chi_{\pi-\delta}$ 
in the low temperature regime: $\chi_{\eta-\delta}\ll \chi_{\pi-\delta}$ for $T<T_{\rm pc}$.
A large discrepancy also shows up in 
the other combination between $\chi_{\pi-\sigma}$ and $\chi_{\eta-\sigma}$: 
$\chi_{\eta-\sigma} \ll  \chi_{\pi-\sigma}$ for $T<T_{\rm pc}$.
Looking at the high temperature regime $T>T_{\rm pc}$, one finds that  
the sizable difference is still kept, 
$\chi_{\eta-\delta}\ll \chi_{\pi-\delta}$, while   
$\chi_{\pi-\delta}$ and $\chi_{\eta-\delta}$
\textcolor{black}{get close to zero}, as shown in the panel (b) of Fig.~\ref{crossover_physical}. 
This tendency is actually consistent with the lattice QCD observation \cite{Buchoff:2013nra,Bhattacharya:2014ara}.
In addition, we also find that  
 $\chi_{\sigma-\eta}$ becomes larger than $\chi_{\pi-\sigma}$ at around $T_{\rm pc}$.  
 As the temperature further increases, $\chi_{\pi-\sigma}$ and $\chi_{\eta-\sigma}$ also approach zero with keeping $|\chi_{\pi-\sigma}| \ll |\chi_{\eta-\sigma}|$ for $T>1.5T_{\rm pc}$.
These trends imply that the chiral symmetry is restored faster than the 
the $U(1)_A$ symmetry in the meson susceptibility functions at the physical value of the current quark masses.

Hereafter, we will vary the current quark masses \textcolor{black}{while keeping the input values of the coupling} constants $g_s$, $g_D$, and the cutoff $\Lambda$, and will investigate the correlations among the susceptibility functions through Eq.~(\ref{WI-def}) as well as the nontrivial  coincidence in Eq.~(\ref{chiral_axial}).
Actually in the present NJL model, 
 as the current quark masses decrease, 
the chiral crossover is changed to the chiral second order phase transition at $m_\pi^{c}\simeq 60 {\rm MeV}$ (corresponding to  $m_l=m_s=1.05\,{\rm MeV}$). Thus \textcolor{black}{the chiral-first order-phase transition domain appears} for $m_\pi< 60 {\rm MeV}$.
In the next subsequent subsections, we will 
%
focus on the chiral crossover and the first order phase transition domains, separately. 










\subsection{Crossover domain}

In this subsection, we evaluate the strange quark mass dependence on the susceptibility functions in the crossover domain.
We allow the strange quark mass to be off the physical value, while the light quark mass is fixed at the physical one, $m_l=5.5$ MeV. 
The present NJL model with this setup exhibits the crossover for the chiral phase transition.

\begin{figure}[H]
\begin{tabular}{cc}
\begin{minipage}{1\hsize}
\begin{center}
    \includegraphics[width=9cm]{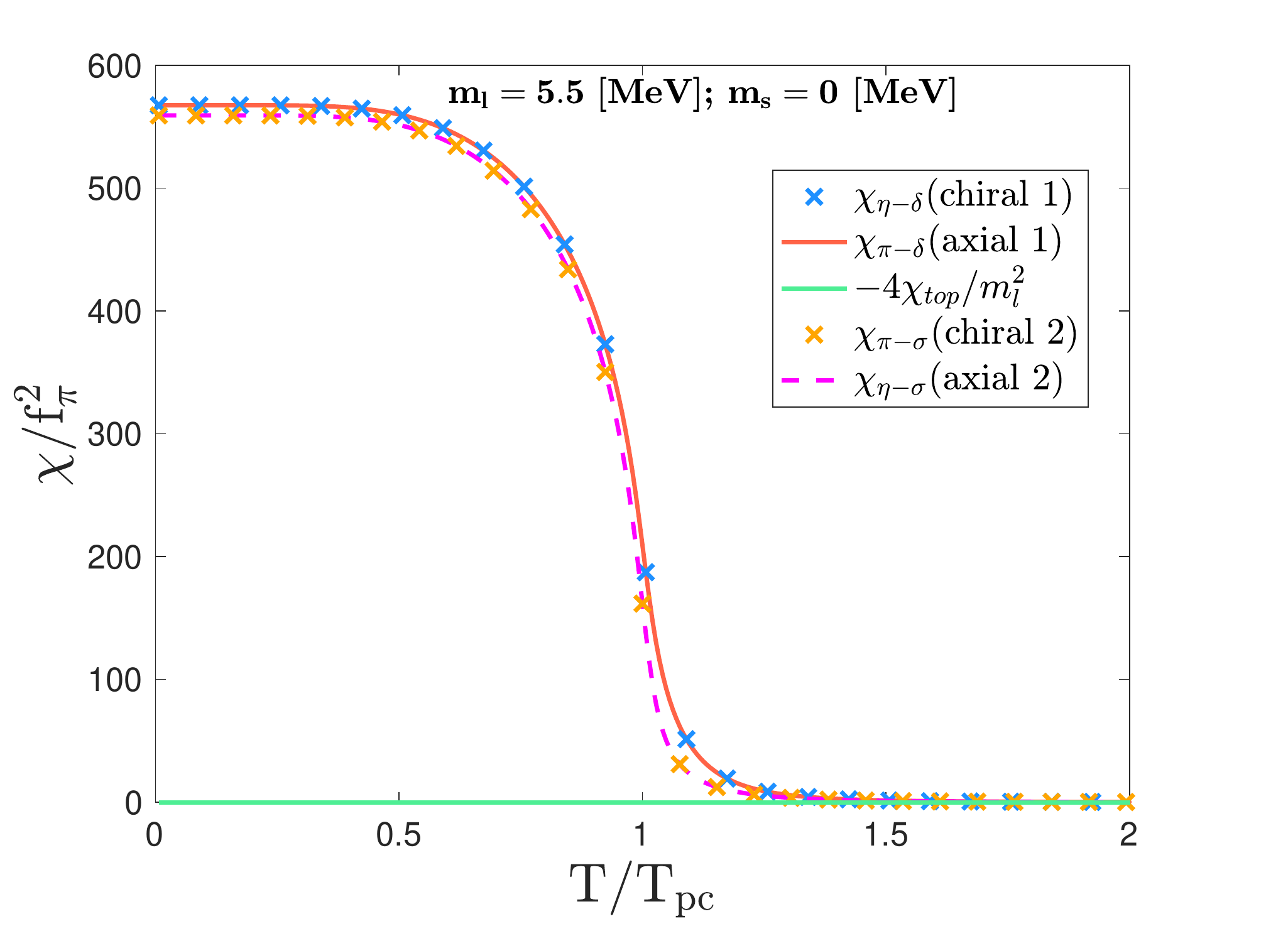}
\end{center}
\end{minipage}
\end{tabular}
\caption{
The plot showing the nontrivial coincidence between the chiral and axial indicators (of two types) in the crossover domain with the massless strange quark ($m_l=5.5$ MeV and $m_s=0$; $T_{\rm pc}\simeq 144$ MeV). 
The topological susceptibility is exactly zero for all temperatures due to the flavor singlet nature associated with the massless strange quark. 
The scaled factors have been applied on both horizontal 
and vertical axes in the same way as in Fig.~\ref{crossover_physical}. 
}
\label{coincidence_CD_ms0}
\end{figure}

In Fig.~\ref{coincidence_CD_ms0}, we plot the susceptibility functions in the massless limit of the strange quark mass ($m_s=0$).
This figure shows that the topological susceptibility $\chi_{\rm top}$ vanishes for any temperature. This must be so  
due to the flavor singlet nature in Eq.~(\ref{FS_condition}).
As the consequence of the vanishing $\chi_{\rm top}$,
the axial indicator $\chi_{\pi-\delta}$ ($\chi_{\eta-\sigma}$) coincides with 
the chiral indicator $\chi_{\eta-\delta}$ ($\chi_{\pi-\sigma}$) for the whole temperature regime, so that the chiral $SU(2)_L\times SU(2)_R$ symmetry is simultaneously restored with the $U(1)_A$ symmetry. 

\begin{figure}[H]
\begin{tabular}{cc}
\begin{minipage}{1\hsize}
\begin{center}
    \includegraphics[width=9cm]{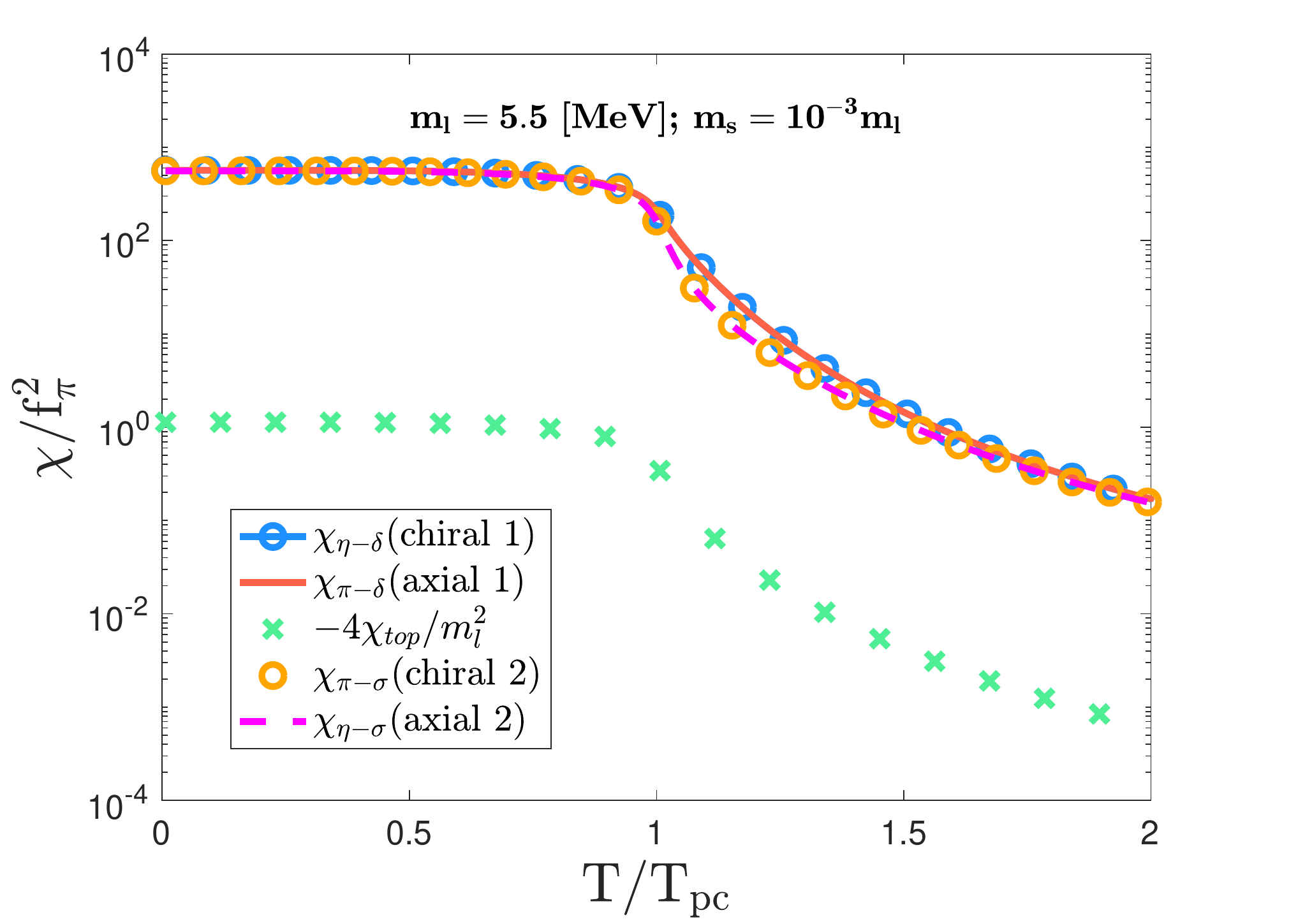}
\end{center}
\end{minipage}
\end{tabular}
\caption{
The plot showing the finiteness of the topological susceptibility along with the temperature dependence of the chiral and axial indicators
in the crossover domain with a small strange quark mass ($m_l=5.5$ MeV and $m_s=10^{-3}m_l$; $T_{\rm pc}\simeq 144$ MeV).
The same scaling for two axes has been made as in Fig.~\ref{crossover_physical}.  
}
\label{mass_CD_small}
\end{figure}

Once a finite strange quark mass is turned on, 
the topological susceptibility $\chi_{\rm top}$ becomes finite, no matter how $m_s$ is small, as seen in Fig.~\ref{mass_CD_small}.  
It is interesting to note that 
when $m_s \ll m_l$, like in Fig.~\ref{mass_CD_small} with 
$m_s=10^{-3}m_l$, 
the topological susceptibility ($-4\chi_{\rm top}/m_l^2$) is much smaller than the chiral indicator $\chi_{\eta-\delta}$ ($\chi_{\pi-\sigma}$) and the axial indicator $\chi_{\pi-\delta}$ ($\chi_{\eta-\sigma}$). 
This is because $\chi_{\rm top}$ is proportional to the $m_s$, as the consequence of the flavor singlet nature in Eq.~(\ref{FS_condition}).
According to the correlation among susceptibility functions in Eq.~(\ref{WI-def}),
the chiral indicator
$\chi_{\eta-\delta}$ ($\chi_{\pi-\sigma}$)  takes 
almost the same trajectory of what the axial indicator $\chi_{\pi-\delta}$ ($\chi_{\eta-\sigma}$) follows at finite temperatures. 
Thus, it is the negligible $\chi_{\rm top}$ that triggers 
the (almost) simultaneous symmetry restoration for the chiral and axial symmetries in the case of the tiny strange quark mass, $m_s\ll m_l$.


\begin{figure}[H]
\begin{tabular}{cc}
\begin{minipage}{0.5\hsize}
\begin{center}
\includegraphics[width=9cm]{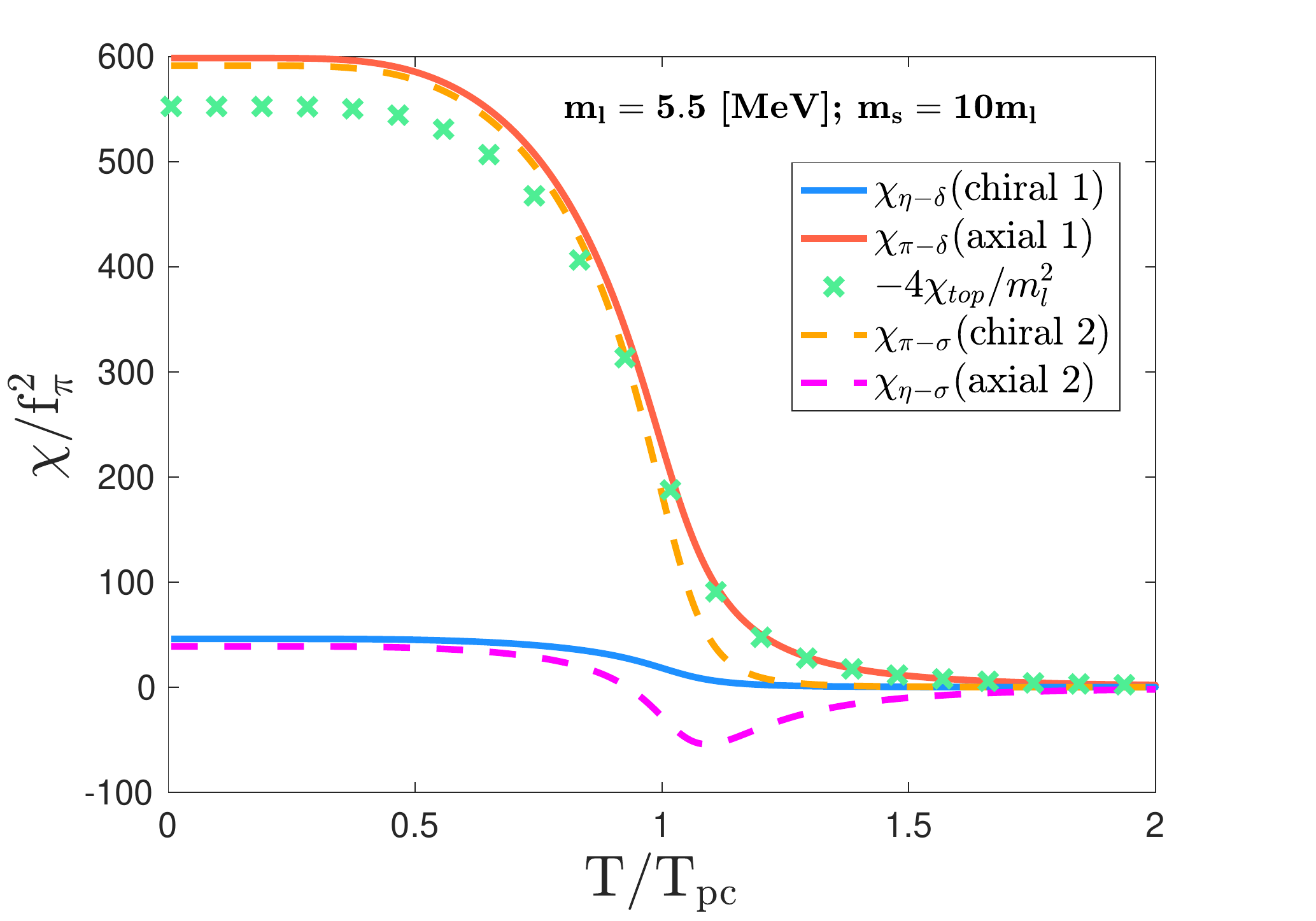}
\subfigure{(a)}
\end{center}
\end{minipage}
\begin{minipage}{0.5\hsize}
\begin{center}
\includegraphics[width=9cm]{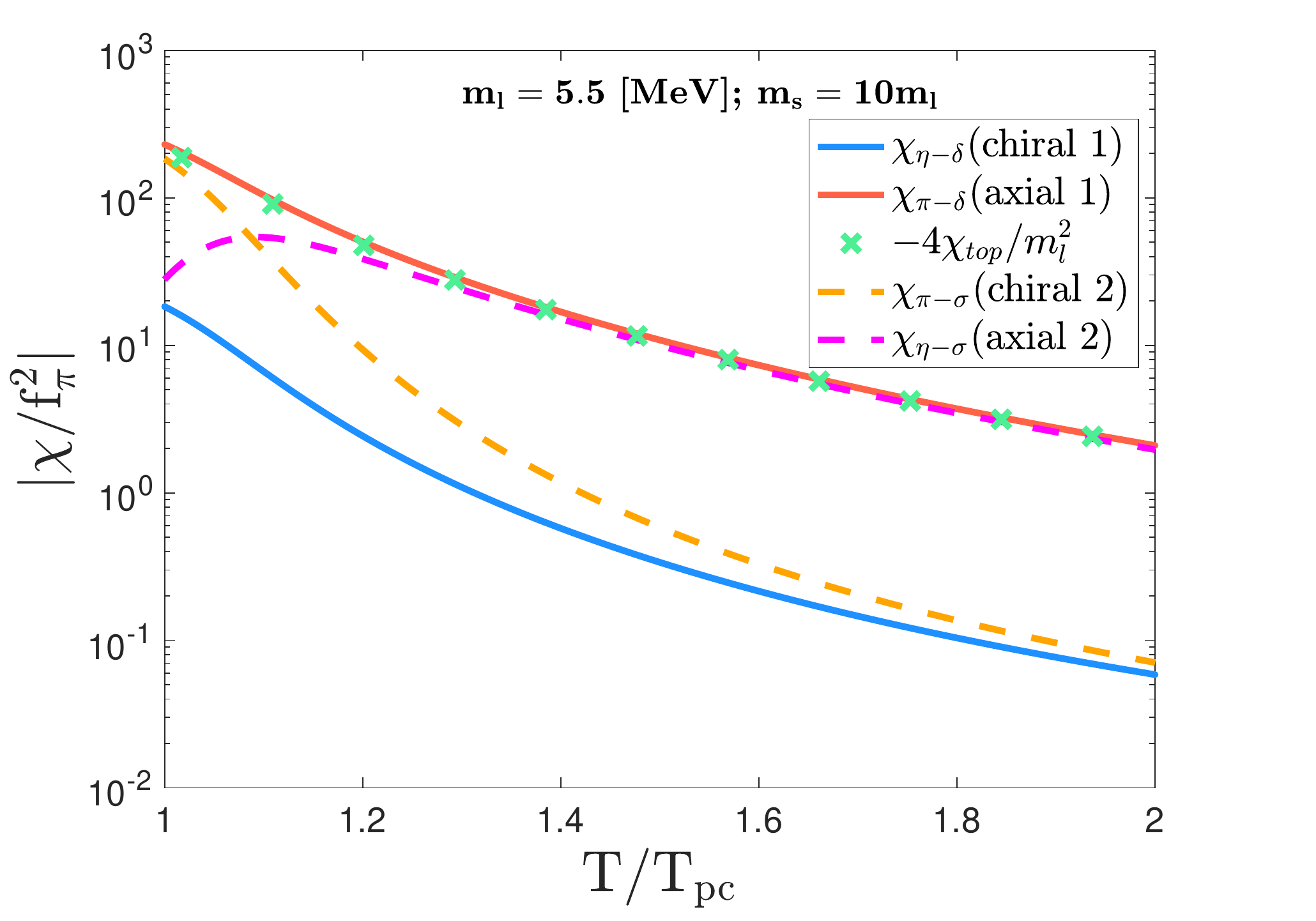}
\subfigure{(b)}
\end{center}
\end{minipage}
\end{tabular}
\caption{
The plots clarifying the significant interference of the topological susceptibility to make the sizable discrepancy between the chiral and axial indicators in the crossover domain with the large strange quark mass ($m_l=5.5$ MeV and $m_s=10 m_l$; $T_{\rm pc}\simeq174$ MeV)  for (a)
$T/T_{\rm pc}=0-2$ 
and
(b) $T/T_{\rm pc}=1-2$.
The manner of scaling axes is the same as in Fig.~\ref{crossover_physical}.  
}
\label{mass_CD_msl}
\end{figure}

As $m_s$ gets greater,
the topological susceptibility further grows and 
$\chi_{\rm top}$ starts to significantly contribute 
to the chiral and axial indicators following the correlation form in Eq.~(\ref{WI-def}). 
Actually, when the strange quark mass takes of 
$O(10\,m_l)
$,
the topological susceptibility ($-4\chi_{\rm top}/m_l^2$) becomes on the same order of magnitude of  
$\chi_{\eta-\delta}$ and $\chi_{\pi-\sigma}$ in the low temperature regime: $ -4\chi_{\rm top}/m_l^2 \sim \chi_{\pi-\delta}\sim \chi_{\pi-\sigma}$ for $T<T_{\rm pc}$.
This trend is depicted in the panel (a) of Fig.~\ref{mass_CD_msl} 
for $m_s=10m_l$.
Thus, the sizable discrepancy between the chiral indicator $\chi_{\eta-\delta}$ ($\chi_{\pi-\sigma}$) and the axial indicator $\chi_{\pi-\delta}$ ($\chi_{\eta-\sigma}$) emerges for $T<T_{\rm pc}$ due to
the interference of $\chi_{\rm top}$.
As the temperature further increases, the susceptibilities go to zero. However  
a sizable discrepancy still appear between the chiral and axial indicators:
$|\chi_{\eta-\delta}|\ll |\chi_{\pi-\delta}|$ and $|\chi_{\pi-\sigma}|\ll |\chi_{\eta-\sigma}|$
for $T>1.5T_{\rm pc}$, as depicted in the panel (b) of Fig.~\ref{mass_CD_msl}. 
This indicates that
the large strange quark mass providing
the significant interference of the topological susceptibility 
urges the faster restoration of the chiral symmetry 
for $m_s =O(10\,m_l)$. 



\subsection{First order domain}
We next consider 
the first-order phase-transition domain.
In this subsection, we fix the light quark mass as $m_l=0.1$ MeV, and vary the strange quark mass $m_s$. 
This setup leads to the first order phase transition
for the chiral symmetry.

\begin{figure}[H]
\begin{tabular}{cc}
\begin{minipage}{0.5\hsize}
\begin{center}
\includegraphics[width=9cm]{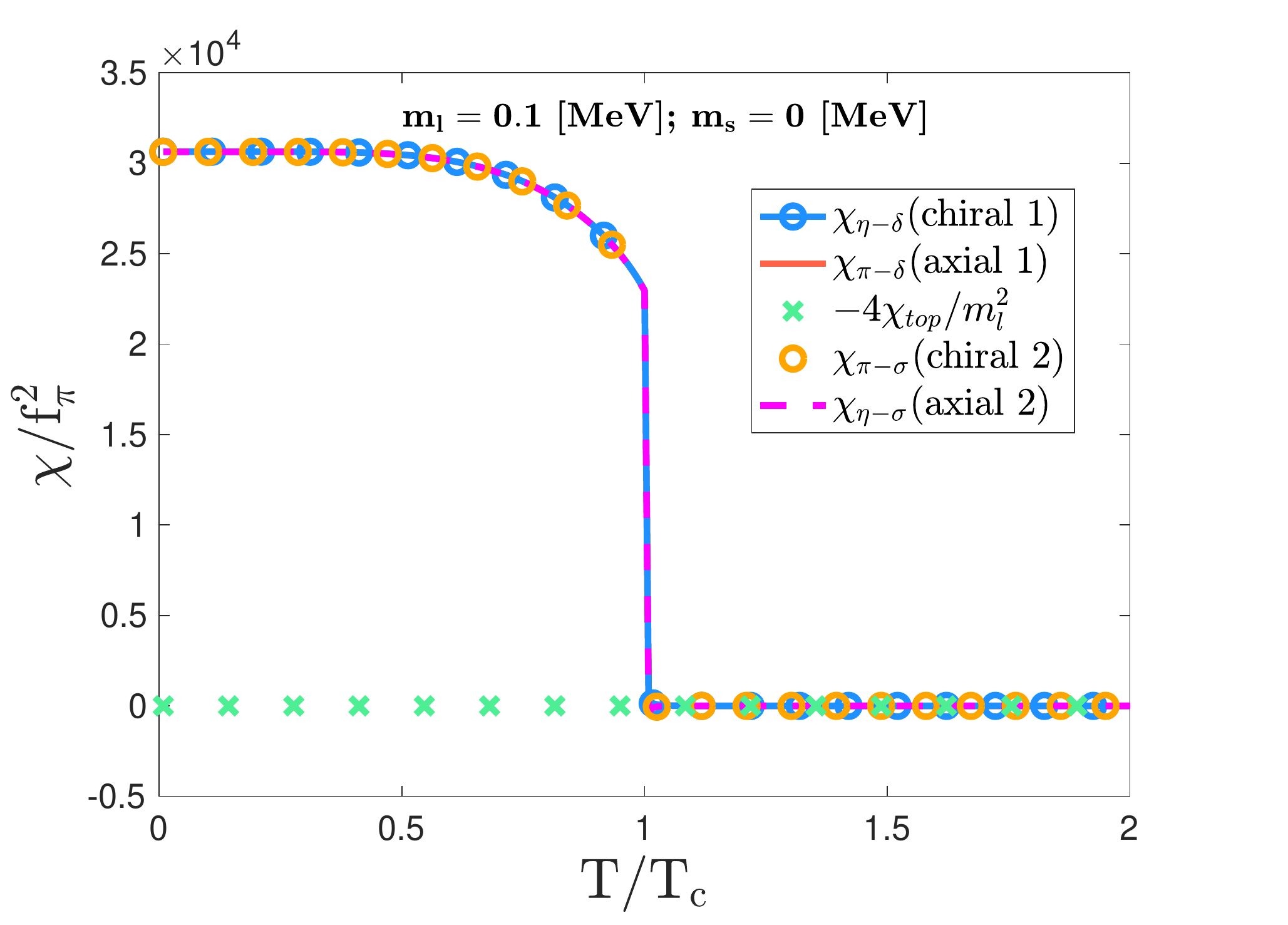}
\subfigure{(a)}
\end{center}
\end{minipage}
\begin{minipage}{0.5\hsize}
\begin{center}
\includegraphics[width=9cm]{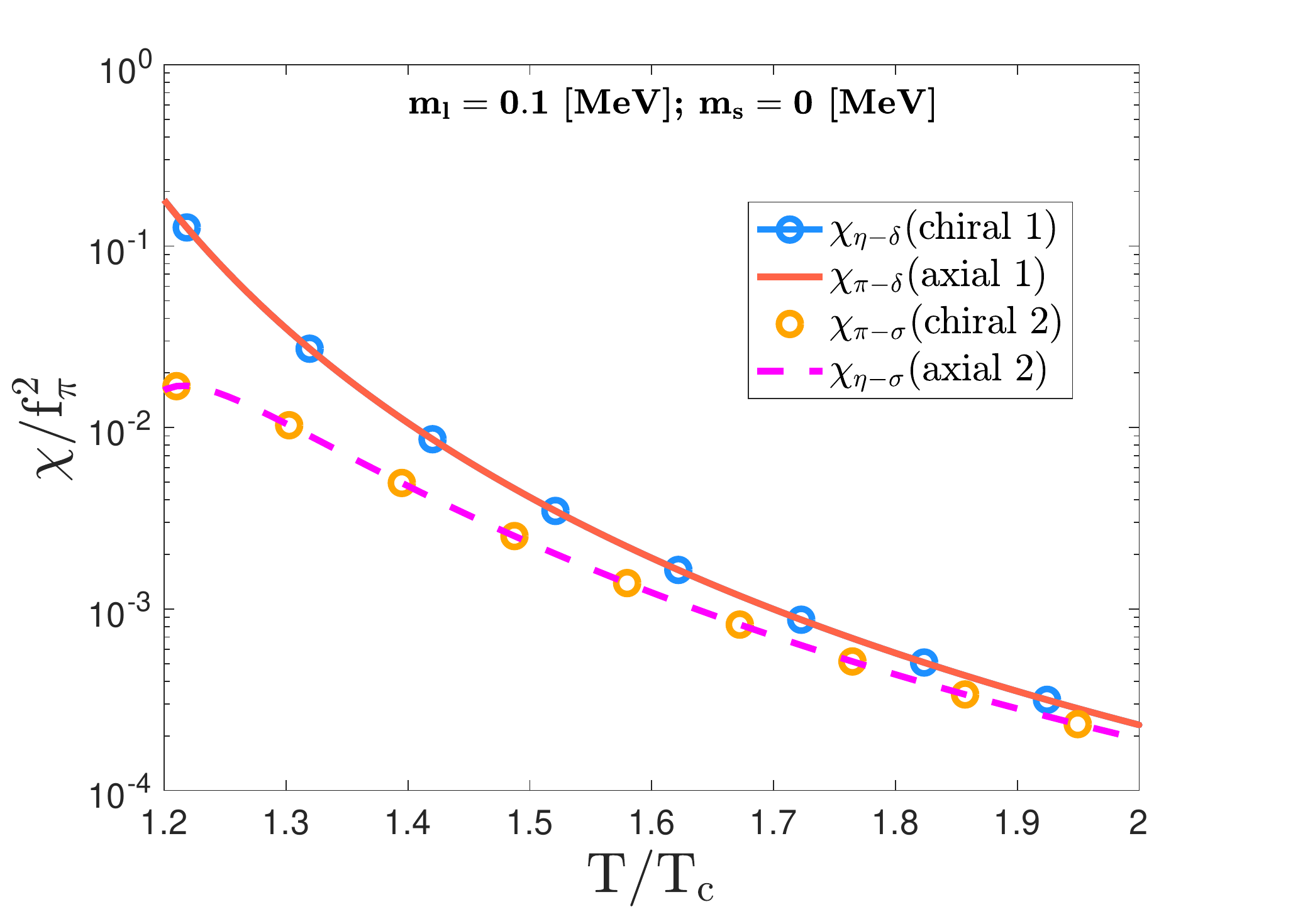}
\subfigure{(b)}
\end{center}
\end{minipage}
\end{tabular}
\caption{
The plots clarifying the trend of the nontrivial simultaneous restoration for the chiral and axial 
symmetries even in 
the chiral-first order phase-transition domain with $m_l=0.1$ MeV and $m_s=0$. 
The panel (a) shows a jump in the mesons susceptibility functions at around  $T_{\rm c}\simeq 119$ MeV, as a consequence of the first order phase transition.
The panel (b) closes up the temperature dependence for the chiral and axial indicators after the chiral phase transition. 
The manner of scaling axes is the same as in Fig.~\ref{crossover_physical}. 
The trend induced by interference of $\chi_{\rm top} $ is similar to the one observed in the crossover domain, in  Fig.~\ref{coincidence_CD_ms0}.  
}
\label{mass_FD_ms0}
\end{figure}

First of all, see Fig.~\ref{mass_FD_ms0}, which shows the susceptibility functions for $m_s=0$. 
One can find a jump in meson susceptibility functions at the critical temperature $T_c\simeq 119$ MeV.
This jump indicates that the chiral-first order phase transition occurs in the NJL model.
Note that even for $T>T_c$ the meson susceptibility functions take finite values
due to the presence of the finite light-quark mass, as shown in the panel (b) of Fig.~\ref{mass_FD_ms0}. This implies that 
the chiral and axial symmetries are not completely restored.  
However, the topological susceptibility is exactly zero because of $m_s=0$, reflecting the flavor singlet nature of $\chi_{\rm top}$. 
This is why we observe  $\chi_{\eta-\delta}=\chi_{\pi-\delta}$ and $\chi_{\pi-\sigma}=\chi_{\eta-\sigma}$ for the whole temperature (see Fig.~\ref{mass_FD_ms0}). 
In particular, 
the panel (b) of Fig.~\ref{mass_FD_ms0} shows that 
$\chi_{\eta-\delta}$ ($\chi_{\pi-\sigma}$) asymptotically goes to zero along with $\chi_{\pi-\delta}$ ($\chi_{\eta-\sigma}$) as the temperature increases.
Thus,
the chiral symmetry tends to simultaneously restore with the axial symmetry 
even in the chiral-first order phase-transition domain.  

\begin{figure}[H]
\begin{tabular}{cc}
\begin{minipage}{0.5\hsize}
\begin{center}
\includegraphics[width=9cm]{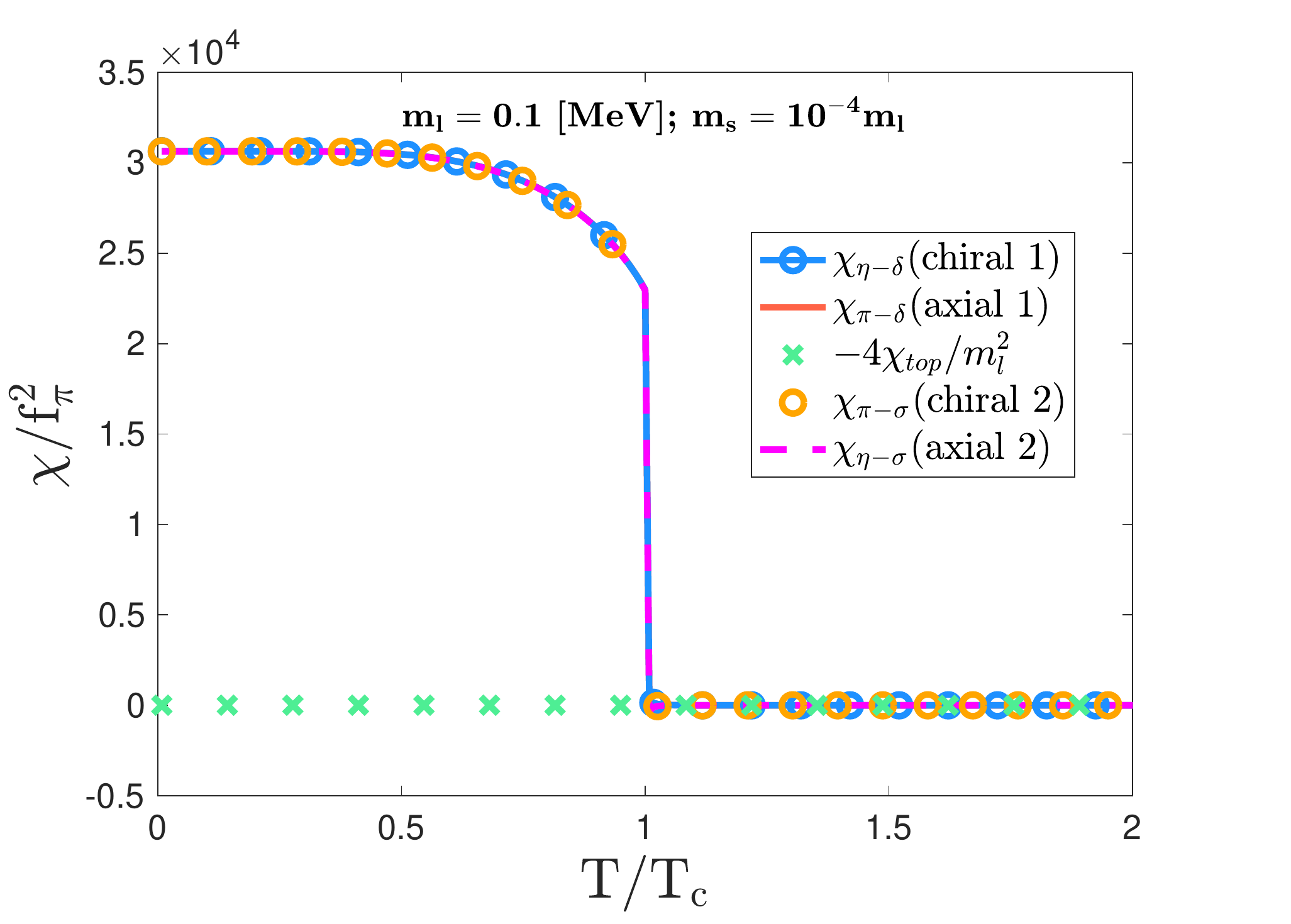}
\end{center}
\end{minipage}
\begin{minipage}{0.5\hsize}
\begin{center}
\includegraphics[width=9cm]{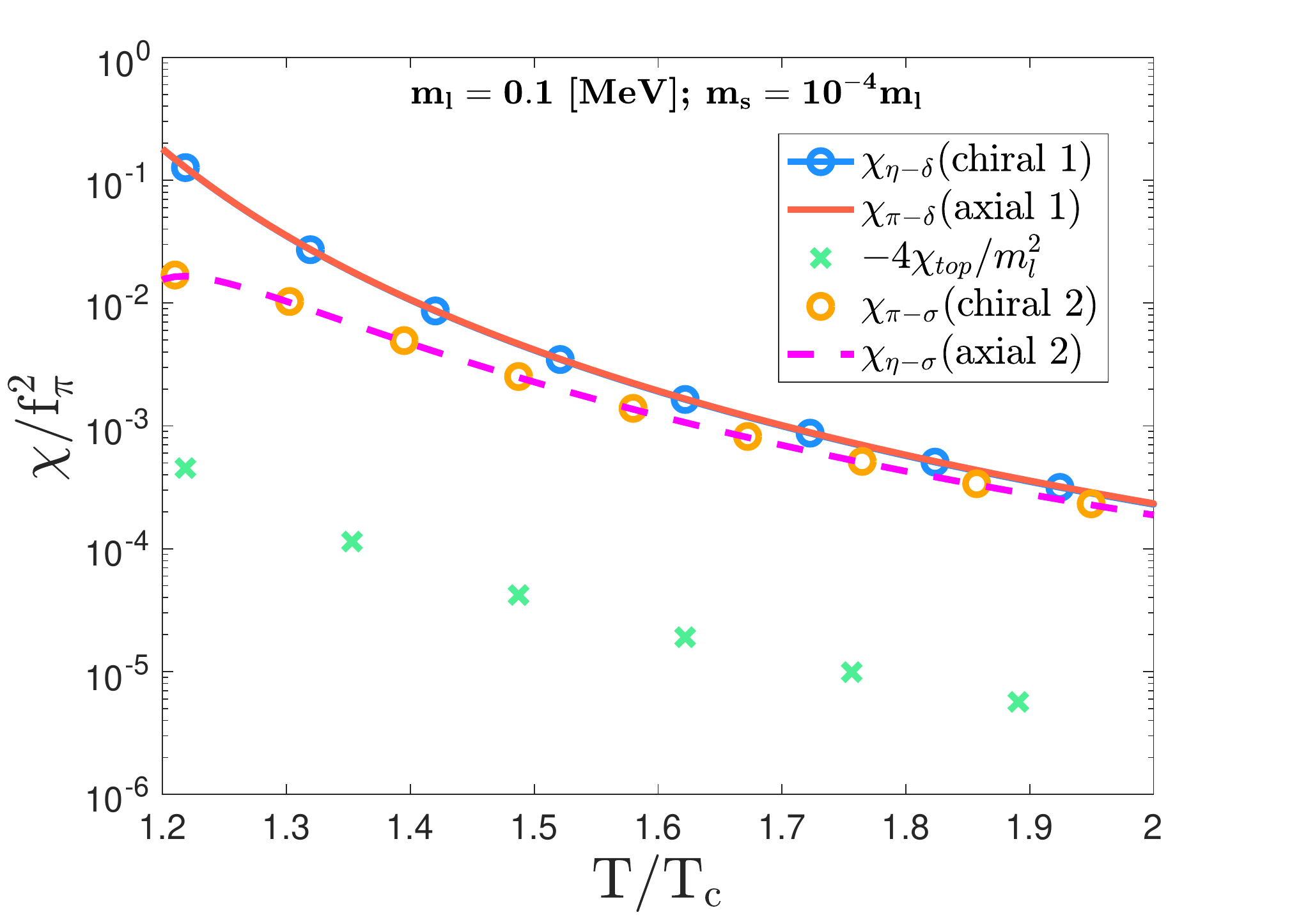}
\end{center}
\end{minipage}
\end{tabular}
\caption{
The plots showing still almost coincidence of the 
chiral and axial indicators for all temperature ranges, 
even in the first-order phase-transition domain with $m_l=0.1$ MeV and $m_s=10^{-4} m_l$; $T_{\rm c}\simeq 119$ MeV. 
The displayed two axes have been scaled in the same way as explained in the caption of Fig.~\ref{crossover_physical}.   The trend induced by interference of $\chi_{\rm top} $ is similar to the one observed in the crossover domain, in  Fig.~\ref{mass_CD_small}. 
}
\label{FD_smallml}
\end{figure}

Next, we supply a finite value for the strange quark mass, 
to see that 
the non-vanishing $\chi_{\rm top}$ certainly emerges in 
the first-order phase-transition domain, as in the case of the chiral crossover domain. 
See Fig.~\ref{FD_smallml}. 
As long as the strange quark mass is small enough,  like $m_s\ll m_l$,  
the topological susceptibility is negligible compared with the chiral and axial indicators.
Therefore, 
the coincidence between the chiral and axial symmetry restoration is effectively almost intact.

\begin{figure}[H]
\begin{tabular}{cc}
\begin{minipage}{0.5\hsize}
\begin{center}
\includegraphics[width=9cm]{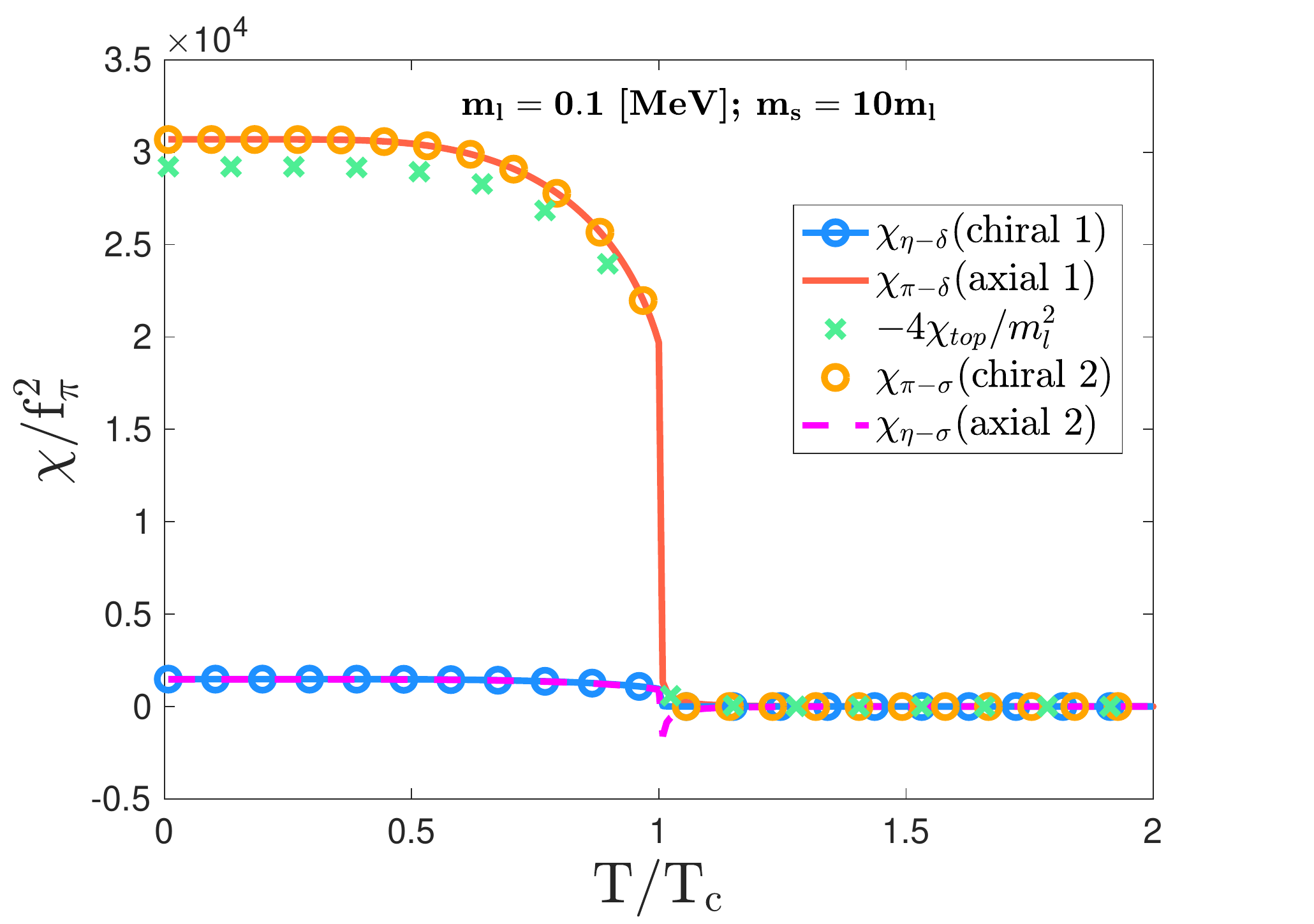}
\subfigure{(a)}
\end{center}
\end{minipage}
\begin{minipage}{0.5\hsize}
\begin{center}
\includegraphics[width=9cm]{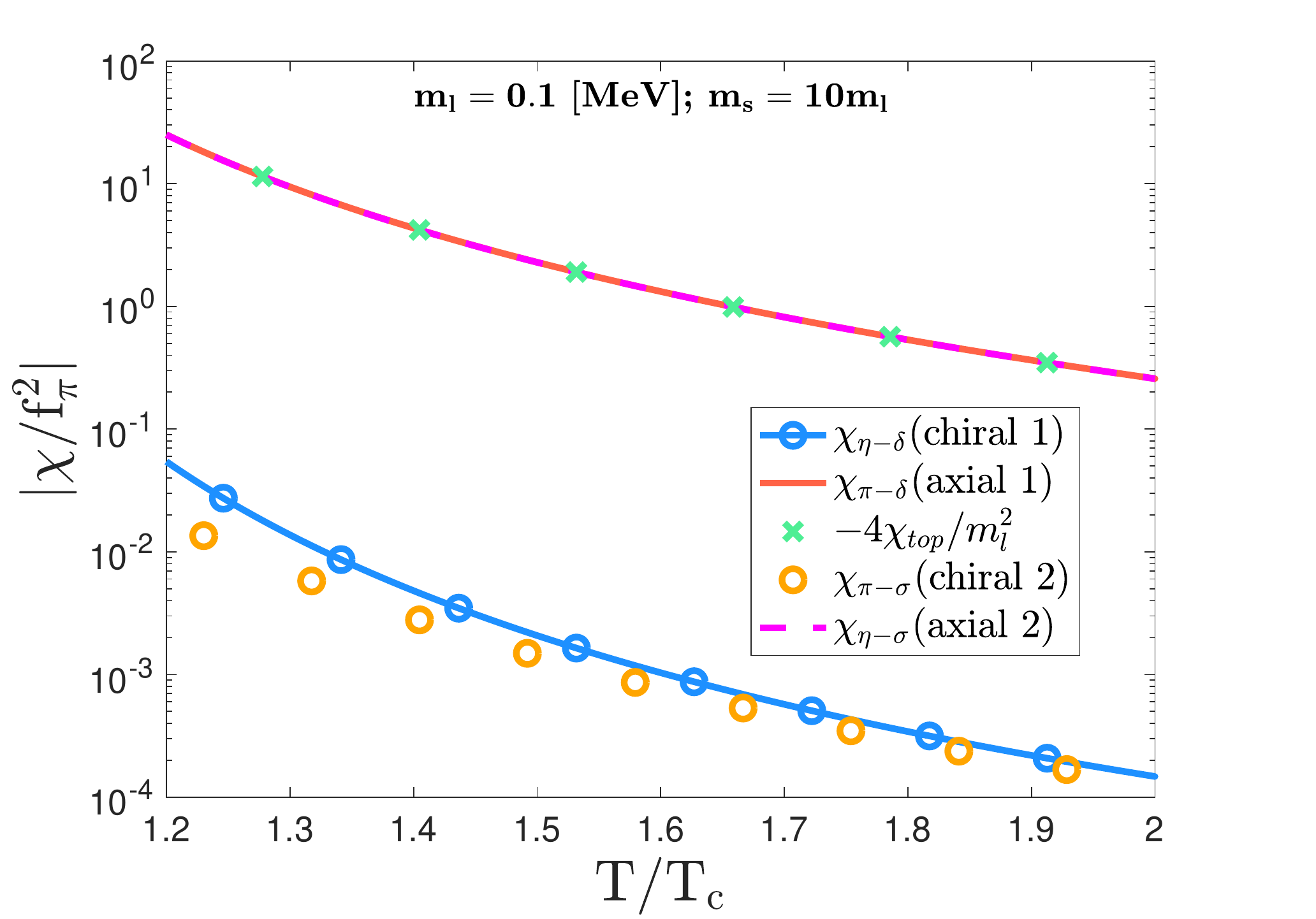}
\subfigure{(b)}
\end{center}
\end{minipage}
\end{tabular}
\caption{
The plots clarifying the significant interference of $\chi_{\rm top}$ with 
the chiral and axial indicators in the first-order phase-transition domain with $m_l=0.1$ MeV and $m_s=10m_l$; $T_{\rm c}\simeq 126$ MeV. 
The displayed two axes have been scaled in the same way as explained in the caption of Fig.~\ref{crossover_physical}.   The trend induced by interference of $\chi_{\rm top} $ is similar to the one observed in the crossover domain, in  Fig.~\ref{mass_CD_msl}. 
}
\label{FD_largeml}
\end{figure}

As the strange quark mass further increases, the topological susceptibility develops to be non-negligible. For $m_s = {\cal O}(10m_l)$, the topological susceptibility ($-4\chi_{\rm top}/m_l^2$) significantly interferes with the chiral and axial indicators via Eq.~(\ref{WI-def}) 
and becomes comparable to $\chi_{\pi-\delta}$ and $\chi_{\pi-\sigma}$ for $T<T_c$ (see Fig.~\ref{FD_largeml}). 
 For $T>T_c$,   
we observe 
a large discrepancy: 
$|\chi_{\eta-\delta}|\ll|\chi_{\pi-\delta}|$ and $|\chi_{\pi-\sigma}|\ll |\chi_{\eta-\sigma}|$, 
due to the significant contribution of the topological susceptibility. 
Thus, the sizable strange quark mass 
makes   
the chiral restoration faster than the axial restoration through the non-\textcolor{black}{negligible} contribution of the topological susceptibility.
Indeed, these trends of the strange quark mass controlling
$\chi_{\rm top}$ in the first-order phase-transition domain are similar to those observed in  the chiral crossover domain.

\subsection{
Chiral and axial symmetry restorations in view of chiral-axial phase diagram }

In the previous subsections, it has been found that 
the topological susceptibility handled by
the strange quark mass is closely related to the meson susceptibilities and interferes with the strengths of the chiral ans axial symmetry breaking.
Here, we clarify more on the strange quark mass dependence on 
the restoration trends of the chiral and axial symmetry. 

\begin{figure}[htbp]
\begin{tabular}{cc}
   \begin{minipage}{0.5\hsize}
     \begin{center}
      \includegraphics[width=9cm]{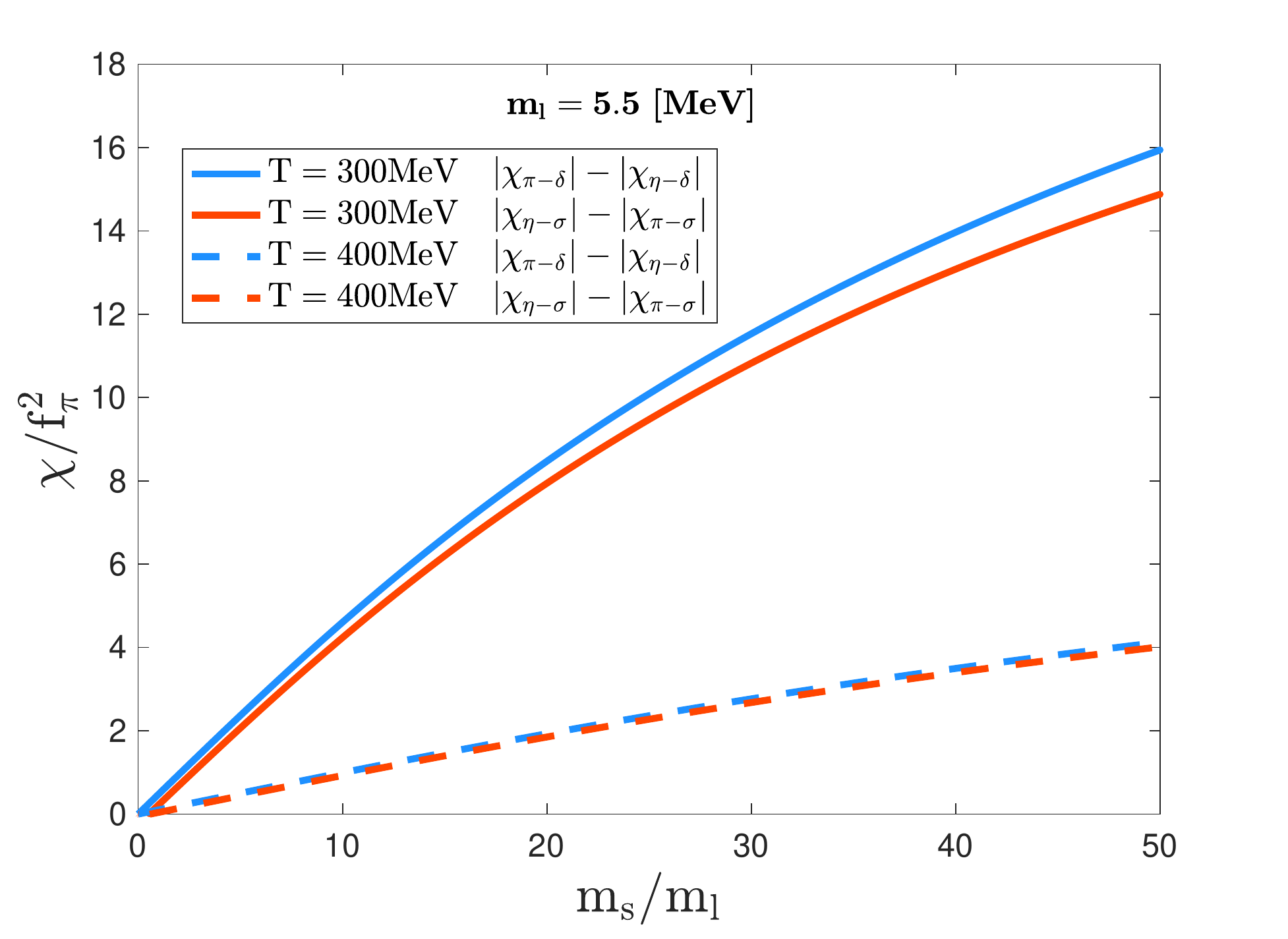}
    \subfigure{(a)}
      \end{center}
    \end{minipage}
    \begin{minipage}{0.5\hsize}
     \begin{center}
     \includegraphics[width=9cm]{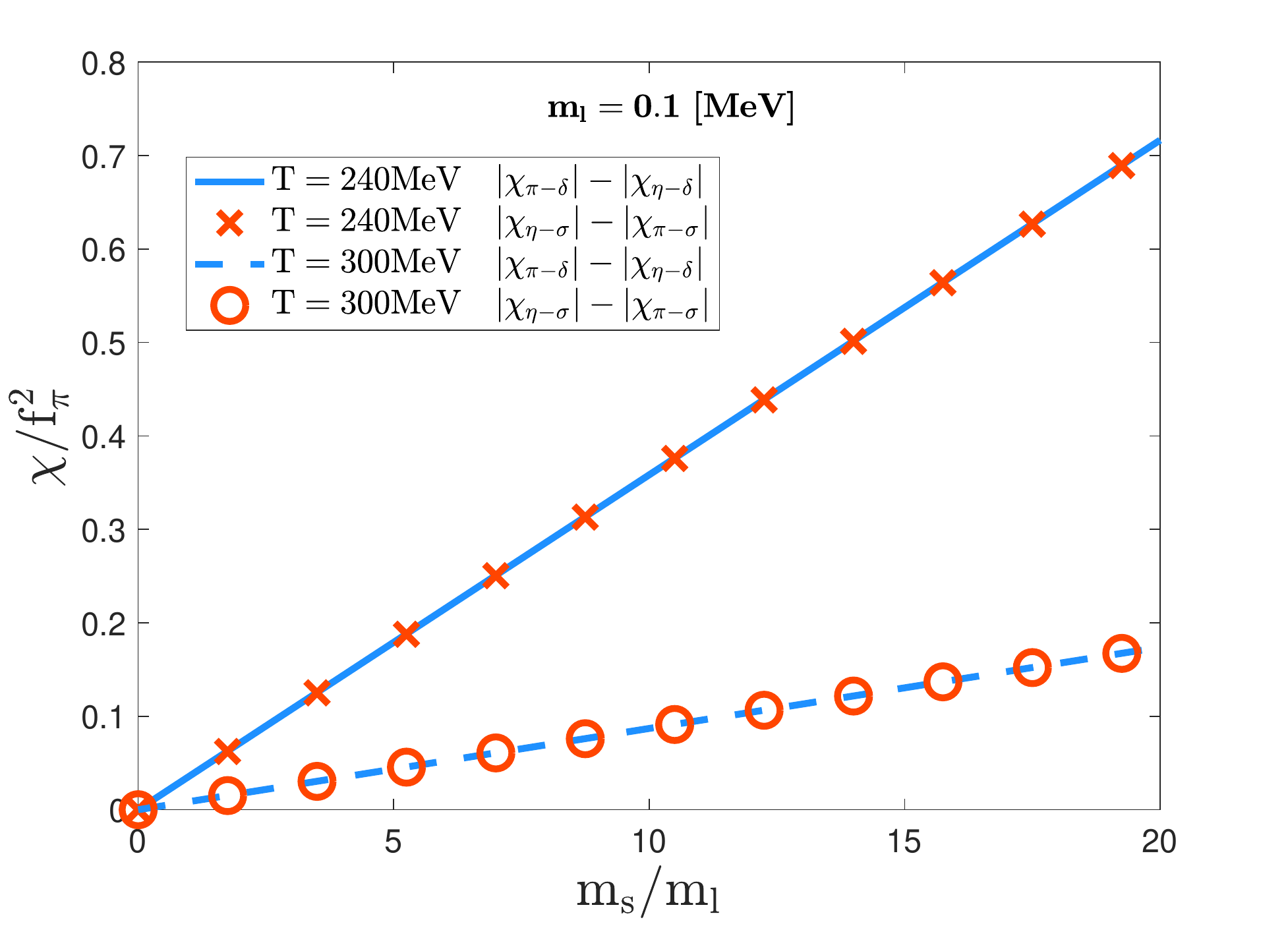}
     \subfigure{(b)}
    \end{center}
\end{minipage}
\end{tabular}
\caption{
The strange quark mass dependence on the difference between the axial indicator $|\chi_{\pi-\delta}|$ ($|\chi_{\eta-\sigma}|$) and
the chiral indicator $|\chi_{\eta-\delta}|$ ($|\chi_{\pi-\sigma}|$)
in  
(a) the crossover domain and 
(b) the first-order phase-transition domain.
In the crossover domain with $m_s/m_l=0\, (50)$, 
the pseudocritical temperature is evaluated as 
$T_{\rm pc}\simeq 144\, (200)$ MeV, and then  
the temperatures $T=300-400$ MeV displayed as in panel (a) correspond to $T\simeq (1.5-2.8)\, T_{\rm pc}$. On the other hand, in the first-order phase-transition domain,
the temperatures  $T=240-300$ MeV as fixed in panel (b) correspond to $T\simeq (1.8-2.5)\, T_{\rm c}$, where 
$T_{\rm c}\simeq 119 \, (130)$ MeV for $m_s/m_l=0\,(20)$. 
}
\label{ms_dep_chiralaxial}
\end{figure}

In Fig.~\ref{ms_dep_chiralaxial},  
we plot the strange quark mass dependence on 
the difference between 
the two indicators, 
$|\chi_{\pi-\delta}|-|\chi_{\eta-\delta}|$ 
and $|\chi_{\eta-\sigma}|- |\chi_{\pi-\sigma}|$,
above the (pseudo)critical temperature $T_{\rm (p)c}$ 
In particular, $|\chi_{\pi-\delta}|=|\chi_{\eta-\delta}|$
and $|\chi_{\eta-\sigma}|= |\chi_{\pi-\sigma}|$ are realized at $m_s=0$ even after reaching a high temperature regime where   
$T\sim (1.5-2.5)\,T_{\rm (p)c}$.
This implies that in the case of the massless strange quark, 
the simultaneous restoration for the chiral and axial symmetries 
takes place in both the chiral crossover and first order phase transition domains.
Once the strange quark obtains a finite mass, the axial indicator
$|\chi_{\pi-\delta}|$ ($|\chi_{\eta-\sigma}|$) starts to 
deviate from the chiral indicator
$|\chi_{\eta-\delta}|$ ($|\chi_{\pi-\sigma}|$) due to the emergence of nonzero $\chi_{\rm top}$.
Actually, Fig.~\ref{ms_dep_chiralaxial} shows that the strength of the axial symmetry breaking in the meson susceptibilities is enhanced by the finite strange quark mass through the interference of the topological susceptibility. 
 Furthermore, the discrepancy between 
the axial indicator
$|\chi_{\pi-\delta}|$ ($|\chi_{\eta-\sigma}|$) and
the chiral indicator
$|\chi_{\eta-\delta}|$ ($|\chi_{\pi-\sigma}|$) persists even at the high temperature $T\sim 2.5\,T_{{\rm p}c}$.
Therefore the axial restoration tends to be delayed, later than the chiral restoration
as the strange quark mass gets heavier.

\begin{figure}[htbp]
\begin{tabular}{cc}
    \includegraphics[width=7.4cm]{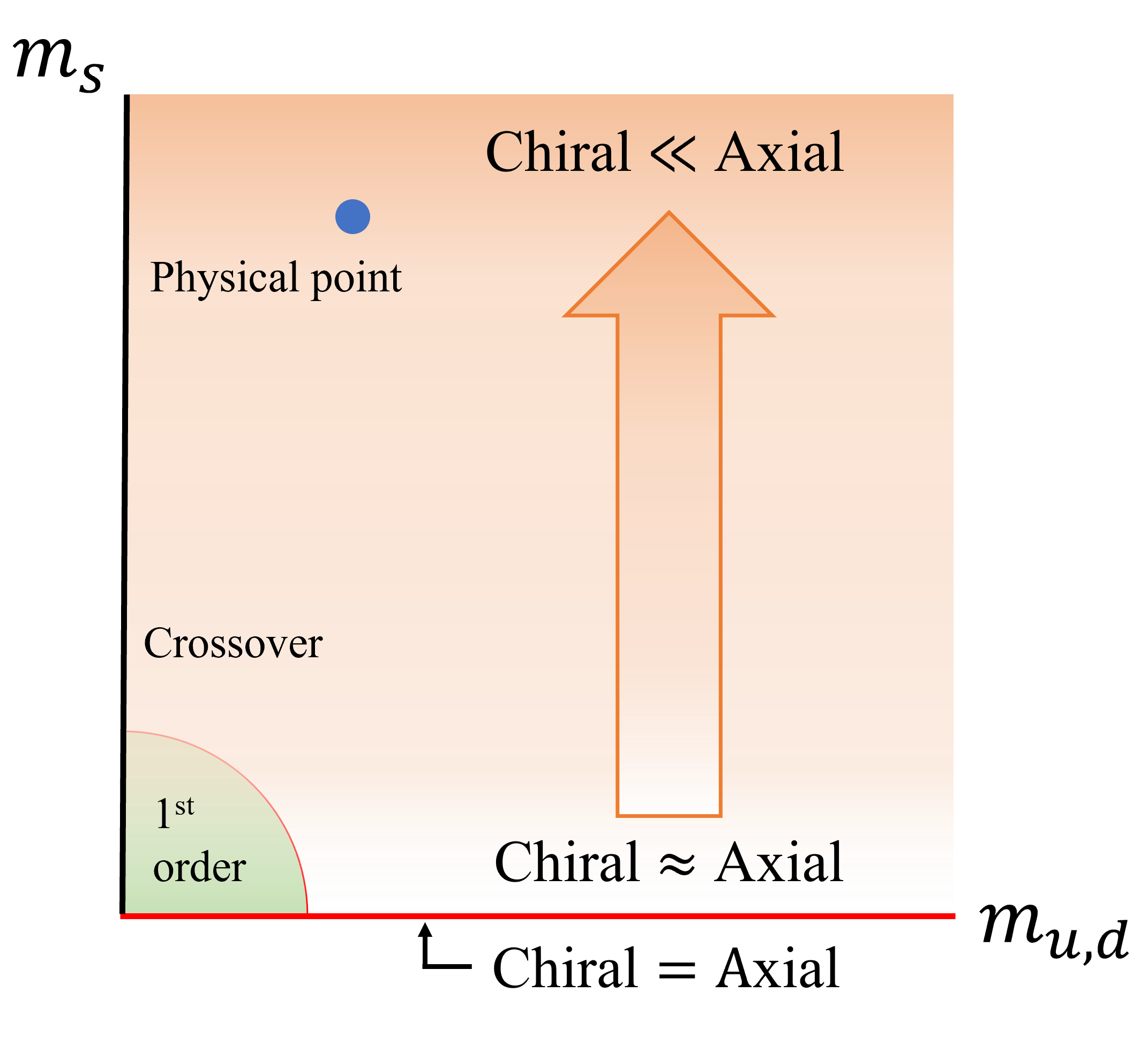}
\end{tabular}
\caption{
The predicted chiral-axial phase diagram on the $m_{u,d}$-$m_s$ plane, in which 
the discrepancy of the chiral and axial symmetry restorations at around $T\sim (1.5-2.0)\,T_{{\rm (p)c}}$ is drawn by the shaded area.
When the strength of the axial symmetry breaking deviates  from the chiral breaking strength to be large, 
the shaded areas become thick. 
The nontrivial coincidence as in Eq.~(\ref{chiral_axial}), 
is 
associated with the vanishing $\chi_{\rm top}$, 
which is located on the $m_{u,d}$ axis.
When the strange quark mass gets a finite mass, the axial restoration deviates from the chiral restoration. 
At around $m_l= O(10m_l)$, the axial restoration is much later than the chiral restoration due to 
the significant interference of $\chi_{\rm top}$.
Namely, at the physical quark masses, 
the topological susceptibility provides the large discrepancy between the chiral and axial restorations in the meson susceptibilities.
}
\label{columbia_NJL}
\end{figure}

Finally, we draw the predicted tendency of the chiral and axial restorations in a chiral-axial phase diagram on the $m_{u,d}$-$m_s$ plane, which is shown in Fig.~\ref{columbia_NJL}.
This phase diagram is a sort of the Columbia plot, in which we reflect the discrepancy between the chiral and axial restorations in terms of the meson susceptibilities at around $T\sim (1.5-2.0)\,T_{{\rm (p)c}}$. 



\section{SUMMARY AND DISCUSSION}

\begin{figure}[htbp]
\centering
   \begin{minipage}[t]{0.5\hsize}
   \centering
      \includegraphics[width=12cm]{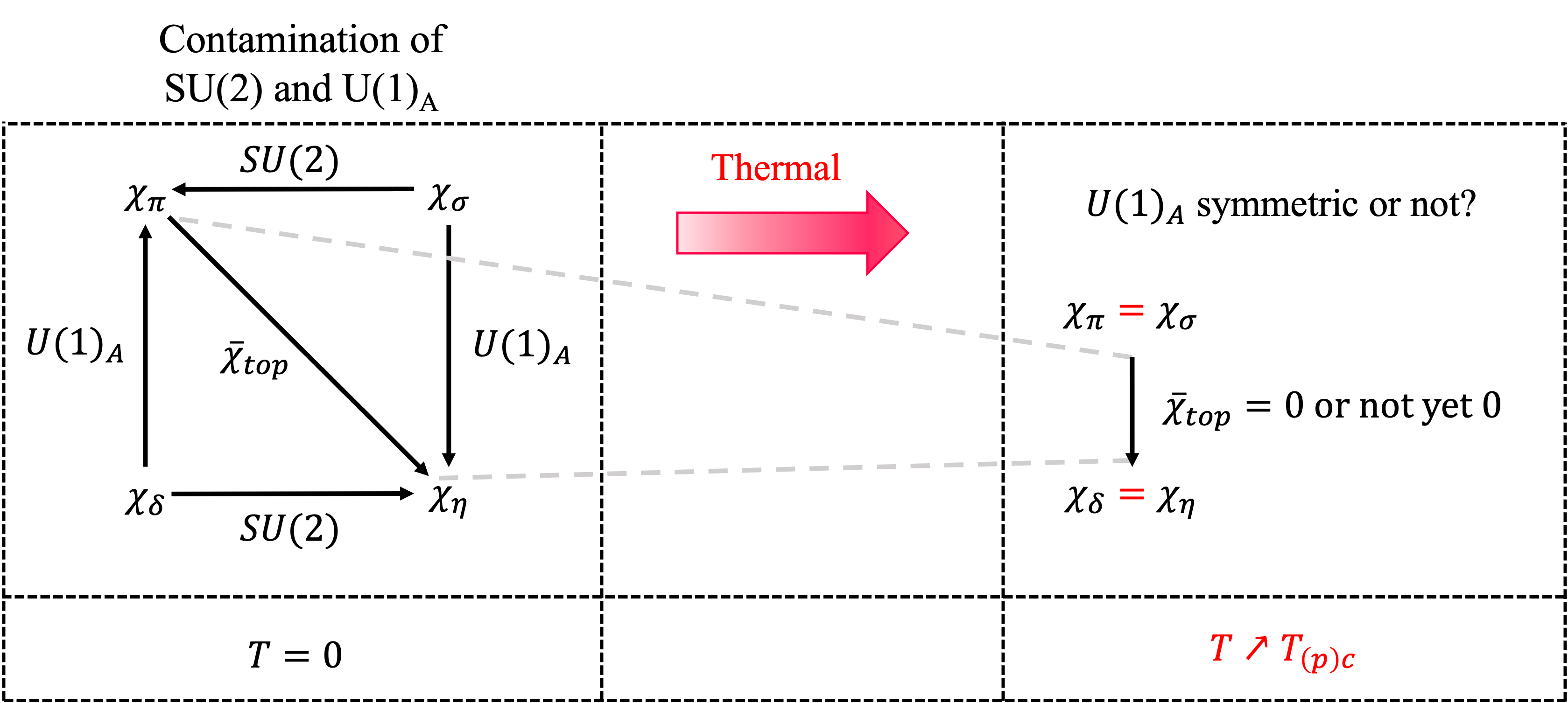}
    \subfigure{(a)}
    \end{minipage}\\
    \begin{minipage}[t]{0.5\hsize}
    \centering
     \includegraphics[width=12cm]{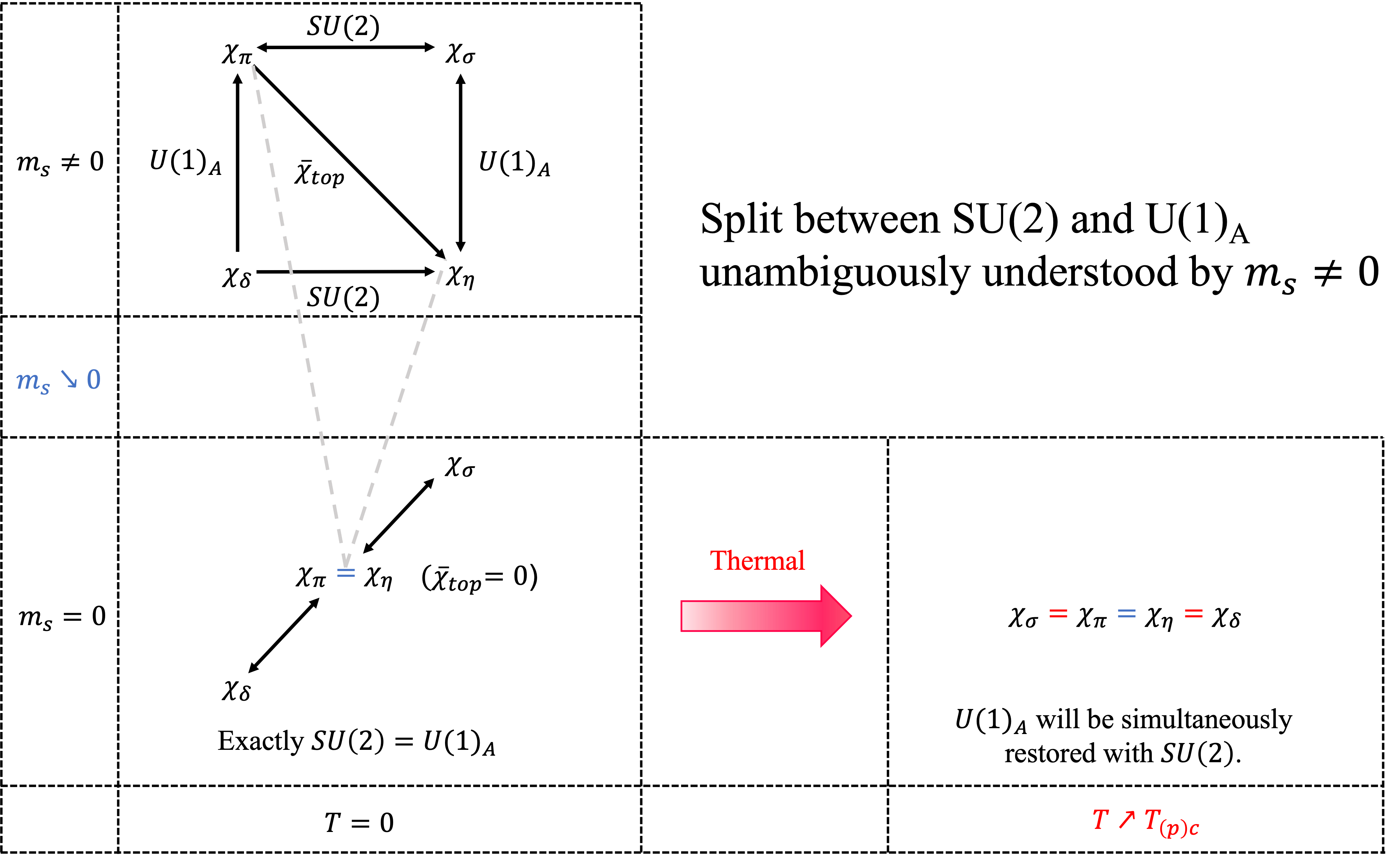}
     \subfigure{(b)}
\end{minipage}
\caption{
The split in the restorations of the chiral $SU(2)$ symmetry and the $U(1)$ axial symmetry at hot QCD. (a): ordinary way to address the symmetry restorations. The ambiguous origin of the effective $U(1)_A$ restoration is often measured by the topological susceptibility (which is normalized by the quark mass, $\bar \chi_{\rm top}=4\chi_{\rm top}/m_l^2$). (b): new point of view for symmetry restorations at $m_s=0$. Due to the anomalous Ward-Takahashi identity at hot QCD, the chiral $SU(2)$ symmetry breaking exactly coincides with the $U(1)_A$ symmetry breaking, and this coincidence holds for any temperatures. As a robust consequence,
when the chiral $SU(2)$ symmetry is restored at the (pseudo)critical temperature, the $U(1)_A$ symmetry is simultaneously restored. Therefore, the limit $m_s=0$ manifests the symmetry restorations on the quark mass plane: it can be unambiguously understood that
the strange quark mass handles 
the split in the restorations of chiral $SU(2)$ symmetry and the $U(1)_A$ symmetry at hot QCD with the three quark flavors having finite masses.}
\label{split_restorations}
\end{figure}

The anomalous chiral-Ward identity in Eq.~(\ref{chitop_wfac}) relating the topological susceptibility to the pseudoscalar meson susceptibility functions has been often used to measure the effective restoration of the $U(1)$ axial symmetry in the $SU(2)$ chiral symmetric phase so far. The $U(1)_A$ restoration probed by the topological susceptibility has extensively been studied in the chiral effective model approaches~\cite{Fukushima:2001hr,Jiang:2012wm,Jiang:2015xqz,Lu:2018ukl,GomezNicola:2019myi,DiVecchia:1980yfw,Hansen:1990yg,Leutwyler:1992yt,Bernard:2012ci,Guo:2015oxa}
and the lattice QCD frameworks~\cite{Buchoff:2013nra,Bhattacharya:2014ara,Borsanyi:2016ksw,Bonati:2018blm,Petreczky:2016vrs} 
to explore the origin of the split in restorations of the chiral $SU(2)$ symmetry and the $U(1)$ axial symmetry.
This ordinary method is summarized in the panel (a) of Fig.~\ref{split_restorations}). 
However, this ordinary approach suffers from practical difficulty 
in accessing the origin of the split 
without ambiguity, because the chiral $SU(2)$ symmetry breaking is highly contaminated 
with the $U(1)$ axial symmetry breaking even at the beginning, at the vacuum.

In this paper,
we have found a new approach: it is the strange quark mass that controls the strengths of
the chiral \textcolor{black}{
$SU(2)$
} symmetry breaking and the $U(1)$ axial symmetry breaking, and those strengths coincide in the limit $m_s=0$. 
The idea is to depart from a nontrivial coincidence limit emerging even in the presence of nonzero $U(1)_A$ anomaly due to the nonperturbatively interacting QCD, in sharp contrast to 
the trivial equivalence between the chiral and axial breaking,  
where QCD gets reduced to the free quark theory. 
Of course, the nontrivial coincidence is robust because it is tied with 
the anomalous chiral Ward-Takahashi identity, Eq.~(\ref{WI-def}), and the flavor-singlet 
condition of the topological susceptibility. 
Hence it holds even at finite temperatures,  
so that the chiral symmetry is simultaneously restored with the axial symmetry regardless of the order of the chiral phase transition. 
The simultaneous restoration at $m_s=0$ is viewed as a significant limit to consider the symmetry restorations on the quark mass plane. 
Given the ``rigid" limit of $m_s \to 0$, 
we can unambiguously understand that the split in the restorations of the chiral $SU(2)$ and $U(1)_A$ is handled by the strange quark mass ($m_s\neq 0$). This new point of view for symmetry restorations is described in the panel (b) of Fig.~\ref{split_restorations}.

To be concrete, 
we have employed the NJL model with three flavors to monitor the essential chiral and axial features that QCD possesses. 
We have confirmed that the NJL model surely provides the nontrivial coincidence of the chiral and axial breaking 
in the case of $m_s=0$ in terms of the meson susceptibility functions,  
and exhibits the simultaneous restoration for the chiral and axial symmetries in both the chiral crossover and the first-order phase-transition cases:     
\begin{eqnarray}
\begin{cases}
\chi_{\eta- \delta}  =   \chi_{\pi- \delta}\to0\\
\chi_{\eta- \sigma}  =   \chi_{\pi- \sigma}\to 0
\end{cases}
, \;\;\;
(\mbox{for }T>  T_{\rm (p)c}, \;\; g_D\neq0,\;\;m_l\neq0\;\;
{\rm and}\;\; m_s=0).
\end{eqnarray}

Once the strange quark gets massive, the topological susceptibility takes a finite value and interferes with the meson susceptibility functions through Eq.~(\ref{WI-def}). 
The simultaneous restoration for the chiral and axial symmetries  
is controlled by  
the strange quark mass through
the interference of nonzero $\chi_{\rm top}$.
Thus, with the large strange quark mass ($m_s \gg m_l$), 
the chiral restoration significantly deviates from the axial restoration above the (pseudo)critical temperature: 
\begin{eqnarray}
\begin{cases}
\chi_{\eta- \delta} \to 0,\;\;    \chi_{\pi- \delta}\to0\;\;\;\mbox{with}\;\;\;
    |\chi_{\pi- \delta}| \gg  |\chi_{\eta- \delta}| \\
\chi_{\eta- \sigma}\to 0,\;\;   \chi_{\pi- \sigma}\to 0\;\;\;\mbox{with}\;\;\;
|\chi_{\eta- \sigma}|  \gg   |\chi_{\pi- \sigma}|
\end{cases}
, \;\;\;
(\mbox{for }T>1.5  T_{\rm (p)c}, \;\; g_D\neq0,\;\;m_l\neq0\;\;
{\rm and}\;\; m_s\neq0).
\end{eqnarray} 
Due to the significant interference of the topological susceptibility,
the chiral symmetry is restored faster than the axial symmetry 
in the $2+1$ flavor case with the physical quark masses. 
Figure \ref{columbia_NJL} shows a schematic view of the evolution of the chiral and 
axial breaking deviating from the nontrivial coincidence limit toward real-life QCD.

In closing the present paper, we give a list of several comments on 
our findings and another intriguing aspect of the nontrivial coincidence between the chiral and axial symmetry breaking strengths. 

\begin{itemize}
\item The predicted chiral-axial phase diagram in Fig.~\ref{columbia_NJL} is a new guideline for exploring the influence of the $U(1)_A$ anomaly on 
the chiral phase transition, which is sort of giving a new interpretation of the conventional Columbia plot. 
Further studies are desired in lattice QCD simulations to 
draw definitely conclusive benchmarks 
on the chiral-axial phase diagram. 

\item  
The existence of the nontrivial coincidence implies that the $U(1)_A$ anomaly can be controlled by the current mass of the strange quark.
The controllable anomaly could give a new understanding of the $\eta'$ meson mass originated from the $U(1)_A$ anomaly. 
The investigation for the 
$m_s$-dependence on the pseudoscalar meson masses 
would thus be a valuable study.

\item 
The fate of the $U(1)_A$ anomaly in the nuclear/quark matter 
is a longstanding problem and has attracted people a lot so far. 
The nontrivial  coincidence should also be realized in the finite dense matter involving the massless strange quark.
The  nontrivial  coincidence at finite density 
might shed light on a novel insight for the partial $U(1)_A$ restoration in the medium with physical quark masses.

\item
The contribution of the $U(1)_A$ anomaly to the color confinement
\textcolor{black}{is an open question.} 
It would be worth including the Polyakov loop terms in the NJL model to address the correlation between the nontrivial coincidence of the chiral and axial breaking and the deconfinement-confinement
phase transition.

\item
\textcolor{black}{Though the NJL model produces the qualitative results consistent with lattice observations, it would be a rough analysis due to the mean field approximation. 
However, the existence of the nontrivial coincidence is robust because it is based on the anomalous Ward identity, 
This should thus be seen even beyond the mean field approximation that the present NJL study has assumed, or even more rigorous nonperturbative analyses such as those based on the lattice NJL-model and the functional renormalization group method. 
}

\item 

The nontrivial chiral-axial coincidence is generic phenomenon, which can also be seen in a generic class of QCD-like theories with ``1 ($m_s=0$)+ 2 ($m_l$) flavors", involving models beyond 
the standard model. In particular, the coincidence in the first-order phase transition case might impact on cosmological implications of QCD-like scenarios with axionlike particles associated with the axial breaking, including the gravitational wave probes. 
Investigation along also this line might be interesting.

\end{itemize}





\section*{Acknowledgements} 

We are grateful to Hen-Tong Ding for useful comments. 
This work was supported in part by the National Science Foundation of China (NSFC) under Grant No.11975108, 12047569, 12147217 
and the Seeds Funding of Jilin University (S.M.).  
The work of A.T. was supported by the RIKEN Special Postdoctoral Researcher program
and partially by JSPS  KAKENHI Grant Number JP20K14479.
M.K. was supported by the Fundamental Research Funds for the Central Universities 
and partially by the National Natural Science Foundation of China (NSFC) Grant  Nos: 12235016,  and the Strategic Priority Research Program of Chinese Academy of Sciences under Grant No XDB34030000.



\end{document}